\documentclass[structabstract]{aa} 
\usepackage{graphicx} 
\usepackage{float}
\usepackage{txfonts}
\usepackage{natbib}
\usepackage{units}
\usepackage{url}
\bibpunct{(}{)}{;}{a}{}{,}
\newcommand{\gro}{GRO J1655-40}
\newcommand{\lsp}{LS~I~+61$^{\circ}$303}
\newcommand{\lsi}{LS~I~+61$^{\circ}$303~}

\newcommand{\beq}{\begin{equation}}
\newcommand{\eneq}{\end{equation}}
\usepackage{color}
\begin{document}

\title{Intrinsic physical properties  and  Doppler boosting effects in \lsp}
\author{
M.\ Massi
 \inst{1}
and G.\ Torricelli-Ciamponi
 \inst{2}
}

\institute{
 Max-Planck-Institut f\"ur Radioastronomie, Auf dem H\"ugel 69,
 D-53121 Bonn, Germany \\
\email{mmassi@mpifr-bonn.mpg.de}
\and 
INAF - Osservatorio Astrofisico di Arcetri, L.go E. Fermi 5,
Firenze, Italy\\
\email{torricel@arcetri.astro.it}
}
\date{Received July 2013; }

\abstract
{}
{Our  aim 
is  to show  how variable  Doppler boosting of an intrinsically variable  jet
 can explain the long-term  modulation of $1667 \pm 8$\,days observed in 
the radio emission of  \lsp.}
{The  physical scenario is that of   
 a conical, magnetized plasma jet having a periodical ($P_1$) increase of relativistic particles, $N_{\rm rel}$, 
at a specific orbital phase, 
as predicted by  accretion  in the eccentric orbit of \lsp.
Jet precession ($P_2$) changes  
the angle, $\eta$, between jet axis and line of sight, thereby 
inducing variable Doppler boosting.
The problem is defined in spherical geometry, 
and  the optical depth through the precessing jet is calculated by taking  into account that the plasma is stratified along the jet axis. The synchrotron emission of such a jet was 
calculated  and 
we fitted the resulting flux density $S_{\rm model}(t)$   to    
the observed flux density  obtained during a 6.5-year  monitoring  of \lsi by the  Green Bank radio interferometer.} 
{Our physical model for the system \lsi  is not only able to reproduce the long-term  modulation 
in the radio emission, but it  also reproduces
all the other observed characteristics of the radio source,
the orbital modulation of the outbursts,  
their orbital phase shift, and their  spectral index properties. 
Moreover, a correspondence seems to exist between
variations in the ejection angle  induced by precession
and the  rapid rotation in position angle observed in  VLBA images. 
}
{The peak of the long-term modulation occurs when the jet electron density is around its maximum  and 
 the approaching jet is forming  the smallest possible angle with the line of sight.
This coincidence of maximum number of emitting particles and  maximum Doppler boosting of their emission
occurs every $\sim$1667 days and creates the long-term modulation observed in \lsp.}


\titlerunning{Doppler Boosting in \lsi}
\maketitle

\section{Introduction}
A recent timing analysis  of the GBI radio data of  \lsi
has revealed  two  frequencies:
$P_1={1\over \nu_1}=\unit[26.49 \pm 0.07]{days}$ and
$P_2={1\over \nu_2}=\unit[26.92 \pm 0.07]{days}$ 
\citep{massijaron13}.
The  period $P_1$ agrees with the value of $\unit[26.4960\pm0.0028]{days}$ \citep{gregory02}
associated to the  orbital period  of the binary system 
formed by a compact object and a Be star 
\citep{grundstrom07}.  
Period  $P_2$  agrees with the previous estimate of 27-28 days  by radio astrometry
of the precessional period of the radio structure 
\citep{massi12}.

The timing analysis results seem to confirm   \lsi  as the  second case of a radio-emitting X-ray binary
with     precessional period ($P_2$) of its  radio jet very close to
the  orbital ($P_1$) period of the binary system \citep{massi12, massijaron13}.
Whereas the several  known precessing X-ray binaries 
\citep{larwood98} have a precession period an order of magnitude
longer than the orbital period (as predicted for  a tidally forced precession by the companion star),  the different case of   GRO J1655-40 was discovered
 in   1995  \citep{hjellmingrupen95}. 
This black hole candidate, which has 
an orbital period of   2.601$\pm 0.027$~days \citep{bailyn95},
revealed a radio jet with a  precessional period  of 
3.0$\pm$0.2 days  \citep{hjellmingrupen95}. 
\lsi is a high-mass X-ray binary, and  \gro{} is a low-mass X-ray binary.
The mechanism proposed in both cases to explain the precession is based on 
general relativity effects around the  
compact object. Lense-Thirring precession has been analysed 
 for \gro{}
by \citet{martin08} and for \lsi{} by \citet{massizimmermann10}.

When  two frequencies are only slightly
different, a beating results,  i.e. an interference 
producing a  new frequency:    their average, 
$ \nu_{\rm average}={{\nu_1+\nu_2}\over 2}$,
which gets modulated with  the beat frequency,
$\nu_{\rm beat}=\nu_1-\nu_2$. As shown  in \citet{massijaron13} 
 for \lsi, the term $1/\nu_{\rm beat}$
 results  $\unit[1667 \pm 393]{days}$ and is compatible with  the observed long-term modulation of the radio flux density 
of $1667 \pm 8$\,days, attributed in the past  
to variations in the wind of the Be star \citep{gregoryneish02}.
Evidence of  beating in \lsi   comes from the fact that the 
periodicity of the radio outbursts
is indeed the one predicted  by the beat theory, which is
$P_{\rm outburst}=P_{\rm average}={1\over\nu_{\rm average}}=26.70\pm 0.05$ days
 \citep{massijaron13, jaronmassi13}. 
Since the the two frequencies 
$ \nu_{\rm average}$ and  $\nu_{\rm beat}$
are related , the long-term modulation  
should also be a result of the beating process between $P_1$ an $P_2$.

The aim of  this paper
is  to prove, in the context of a physical scenario,
 how the mutual relationship between 
$P_1$ and $P_2$ can generate the long-term periodicity.
We linked the two periodicities, $P_1$ and $P_2$, to two physical
processes: the periodical ($P_1$) increase  of relativistic electrons
in a conical  jet
and  the periodical ($P_2$)  Doppler boosting of the
emitted radiation by relativistic electrons because of jet precession.
The hypothesis that  the compact object in \lsp, accreting  material from the  Be wind,
undergoes a  periodical  ($P_1$) increase in the accretion  at a  particular orbital phase
along an eccentric orbit
was suggested and developed  by several authors  \citep{taylor92, martiparedes95, boschramon06, romero07}.
The hypothesis that a
precessing jet, with an approaching jet having  large excursions in its position angle,
should give rise to appreciable Doppler boosting effects
 is supported by the morphology of \citet{massi04}, \citet{dhawan06}, and \citet{massi12}
 images  showing extended radio structures changing from two-sided to one-sided morphologies
at  different position angles.

Our model describes a precessing ($P_2$) jet with a periodically ($P_1$) varying number of
relativistic electrons.
Section~2 describes our analytical model and how to calculate 
its flux density $S_{\rm model}$.
In Sect. 3,   
all parameters  for calculating  $S_{\rm model}$ 
are derived by fitting the model to  6.5 years of GBI radio flux density observations. 
In Sect.\,4 we present our results and in Sect.\,5 our conclusions.

 \begin{figure}[ht]
   \centering
  \includegraphics[scale=.3, clip]{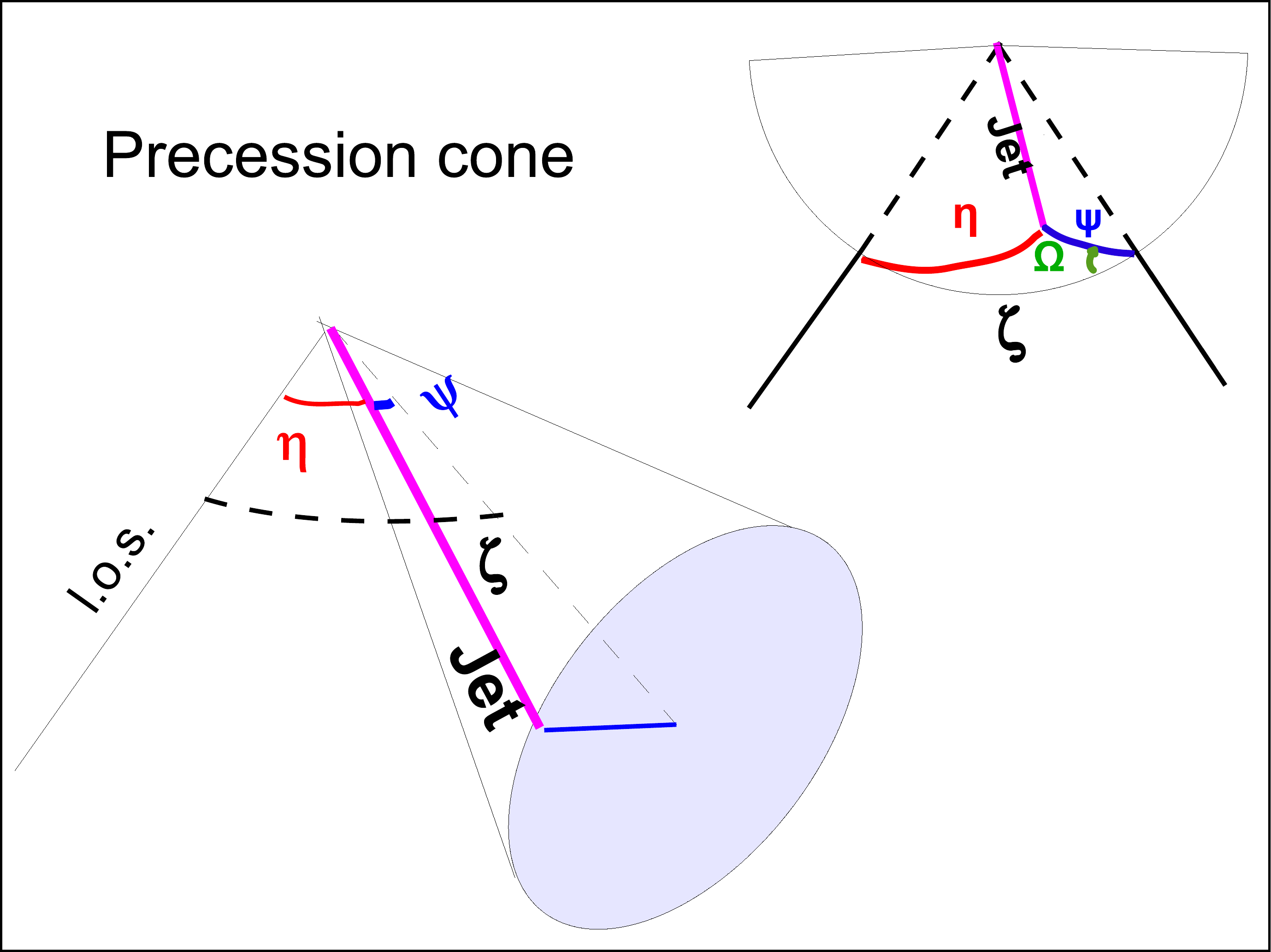}\\
   \caption{Precessing jet and relative angles. Left: Sketch of the precessing cone.
The angle $\zeta$ is the angle between the jet precession axis 
and the line of sight (l.o.s.)  direction,
  $\psi$ is the opening  angle of the precession cone and 
 $\eta$ is the continuosly changing angle between  the jet and the line of sight because of precession.
Right: 
The  angles $\zeta$,$\eta$,$\psi$ in spherical geometry.
  The angle $\Omega$, describing in
  the precession period $P_2$ an angle  $2\pi$,  defines the temporal dependence.  
}
   \label{Fig1}
\end{figure}

\section{Model definition}

In this section we want to analytically determine  the flux density of a precessing 
radio jet having a periodical variation in the number of its emitting particles.
Section 2.1 describes the type  of emitting source 
we are assuming  for our model, which is 
the  conical jet used for microquasars and AGNs after 
the seminal paper of \citet{blandfordkoenigl79}. 
In particular, we adopt  Kaiser's analytical treatment \citep{kaiser06}
to express the changes of magnetic field and  plasma
properties along the jet axis.
The flux density of our jet, which is $S_{\rm model}$ in Eq. 1,
contains  two  contributions: the Doppler factors and the intrinsic 
synchrotron emission of the conical jet.
The Doppler factors are determined in Sect.\,2.2.

The most important parameter introduced  to define the Doppler boosting term
is the  angle $\eta$ between jet axis and line of sight. This is the angle
that continuously varies because of precession.
Section 2.3 deals with the second contribution of Eq. 1, i.e.,
 the intrinsic synchrotron emission of the conical jet.
The difference with   respect to   
Kaiser's  model, where  
the jet is seen at  $\eta \sim 90^{o}$, is that  
here we are dealing  with a precessing jet, i.e., with a continuously varying
angle $\eta$. 
This  implies  the calculation of the optical depth  through a stratified jet (Sect. 2.3 and Appendix).
Also, in our case,  the number  
 of relativistic particles, $N_{\rm rel}$,
changes along the orbit of \lsp, as  described in Sect.~2.4. 

\subsection{The model}

The general framework
for  \lsi emission in the radio  energy range is that of synchrotron radiation emitted  
from a relativistic electron population  of density $N_{\rm rel}$  moving  in a  magnetic field $B$. 
However, this simple scenario does not reproduce the radio spectra
of \lsp.
A  simple magnetized plasma containing particles 
with a power-law energy distribution produces a power-law 
spectrum  ($S \propto \nu^{\alpha}$) with spectral slope $\alpha < 0$ above the  critical frequency,
while 
below this critical frequency   self-absorption effects become important and
 the emission becomes optically 
thick with a spectral slope  $\alpha$=2.5 \citep{longair,kaiser06}.
This spectrum of a uniform, self-absorbed synchrotron source is not
the kind of radio spectrum observed in \lsp. 
The optically thick emission observed in  \lsi is much flatter with 
$\alpha$ in the range of $0.0-0.5$ \citep{massikaufman09, zimmermann13}, similar to
 the values  $\alpha \sim 0.0-0.6$ observed in microquasars \citep{fender01}. 
Optically thick synchrotron emission in microquasars and  in radio loud AGN 
is  modelled with conical jets where changes in  
magnetic field  and   plasma density  along the flow  give rise to 
the observed flat spectra  \citep{blandfordkoenigl79, kaiser06}.
 
In the framework of this kind of jet model, described by unconstant physical quantities,
we follow  Kaiser's model for adiabatic jets \citep{kaiser06},  
to which one can refer for  more details.  
We use his notation when  not otherwise  specified. We therefore assume
 a conical jet, where  all quantities are radially 
constant and change only along the jet axis $x$, with  $x$  expressed 
as $x =x_0 l$,  where $x_0$ is  an arbitrary position along the $x$-axis
defining the dimensionless coordinate $l$
(see Fig.~A1 in Appendix). The cone radius is $r(l)=r_0 l^{a_1}$, and we choose $a_1=1$ to have a  cone with a  constant opening angle $\xi$ (in Kaiser's notation $\tan \xi = r_0/x_0$).

The evolution of the strength of the magnetic field along the jet 
is parametrized as in \citet{kaiser06}  as $B=B_0 l^{-a_2}$.
Depending  on the  magnetic field geometry,  
a purely parallel 
 magnetic field $B \equiv B_{\parallel}$
to the jet axis or a  purely perpendicular 
magnetic field
$B \equiv B_{\perp}$ 
to the jet axis
 imply  different values for
 $a_2$.
From Kaiser's Table 1 we have  for  $B\sim B_{\parallel}$,  $a_2=2$,
while for $B\sim B_{\perp}$,   $a_2=1$. 
As a matter of fact,   
the most likely configuration is 
a helical  magnetic field configuration with  
both the parallel and the perpendicular components.
Therefore, in the procedure described in Sect. 3 
we have calculated the jet emission in both of these limiting cases. 

The number density of the relativistic electrons has the usual
power-law   dependency on the  electron energy:   
$N_{\rm rel}=\kappa  E^{-p}$.  
As in \citet{kaiser06}, we represent 
the evolution of the electron distribution density along the jet by setting 
$\kappa=\kappa_0 l^{-a_3}$, with $a_3=  2(2+p)/3$ and  $p=1.8$ (see Sect. 4.1).
But   in our case 
we also introduce a temporal dependence for electron density,  setting  $\kappa_0$ as  function of 
time with periodicity $P_1$ (see Sect. 2.4).

In this framework we express the total flux density of the two jets,
the approaching, and the receding jet, as
\beq
S_{\rm model}(t)=S_{\rm a}(t) (\delta_{\rm a}(t))^{k-\alpha}+S_{\rm r}(t) ( \delta_{\rm r}(t))^{k-\alpha}
\eneq
where  $\delta_{\rm i}$ is the Doppler boosting term,   with i=a for the approaching jet 
and i=r, for the receding jet. The  emission term, $S_{\rm i}(t)$, similarly represents  
the  emission of each jet.
Parameter  $k$  accounts for the  ejecta properties, with $k$=2 for a continuous jet (used here)
 and $k=3$ for discrete condensations; $\alpha$ is the spectral index. 

The quantities $S_{\rm i}$ and $\delta_{\rm i}$ depend  on the two {\it physical} periodicities $P_1$, the orbital period,  and $P_2$, the precessional period, as described in the following.

\subsection{The Doppler boosting effect and  its dependency on $P_2$}
Synchrotron radiation from relativistic electrons spiralling in
the  magnetic field of the approaching jet  
becomes enhanced  by the Doppler effect (i.e., "boosted"), while 
the radiation from electrons of the
receding jet  becomes
attenuated (i.e., "de-boosted"). 
Doppler boosting and de-boosting strongly depend on the angle between the jet  and the observer's line of sight
\citep{urrypadovani95,mirabelrodriguez99}.
A precession of the jet implies a variation in this angle and therefore
a variable Doppler boosting.

The Doppler boosting and de-boosting terms $\delta_{\rm i} (t)$ are defined as
\beq
\delta_{\rm a}= {\sqrt{1-\beta^2} \over 1-\beta~\cos \eta} 
\eneq
\beq
\delta_{\rm r}= {\sqrt{1-\beta^2} \over 1+\beta~\cos \eta}
\eneq
where  $\beta= {\rm v}/c$ 
(with ${\rm v}$  the  electron velocity component along the jet axis).

Angle $\eta$ is  the angle between
observer's  line of sight and the jet axis. To follow its  changes with time due to the jet precession we express  $\eta$ in terms of other quantities, throughout the   spherical triangle of sides  $\zeta, \psi,$ and $\eta$
 shown in Fig.\ref{Fig1}. 
Angle $\zeta$ is the angle the jet precession axis makes  
with the  direction of the line of sight.
If the jet precession axis  is tilted by an angle $f$ with respect to 
the orbital axis of the  binary system, then $\zeta=i+f$, where $i$ is the system inclination angle;
$\psi$ is the opening  angle of the cone  described by the jet during 
its precession period, $P_{2}$.  
In the spherical triangle of Fig.\,1, the angle $\Omega,$  
between sides $\psi$ and $\zeta$, changes with time because of precession 
 and is defined as 
\beq
\Omega = 2 \pi (t-t_0+\Delta)/P_{2},
\eneq
where 
  $t_0$=JD2443366.775 \citep[and references therein]{gregory02}, and   $\Delta$ 
is the offset to be determined in our model to synchronize $P_2$ to $P_1$.

 From spherical geometry \citep[Eq. 8]{smartgreen77} it follows 
that the time (i.e.  $\Omega$) dependency of  the angle  $\eta$ is
 \beq
 \cos ~\eta = \cos~ \zeta  \cos ~\psi + \sin ~\zeta \sin ~\psi  \cos ~\Omega.
 \eneq
 Inserting the $\eta$ expression into  Eqs. (2) and  (3), we   get  a temporal modulation of the Doppler boosting factor  and we introduce in the  observed flux density the dependency on the precession period $P_2$.

\subsection{Intrinsic jet emission}

In our precessing conical jet model, angle $\eta$, between the jet axis and the line of sight,
 must be clearly variable and can assume all values in the range 
$90^o > \eta > \xi $, with $ \xi$ the jet opening angle.
This is an important difference with   respect to   
Kaiser's  model, where  the jet is seen almost perpendicular to its axis  (i.e. $\eta \sim 90^o$), or to
\citet{pottercotter12}'s  model  where the jet is observed at an angle smaller than 
the opening angle of the jet (i.e., $\eta \le \xi $).

Since in our model we  allow 
 time-dependent  values for angle $\eta$,  the plasma optical depth  also  will depend on time. In fact,
 the optical depth must be computed across the jet along a path that makes an angle $\eta$ with the jet axis and  intersects regions characterized by different values of physical quantities (since in our model all quantities vary along the jet axis, i.e., along $x=x_0l$) (Fig. A1-Bottom). 
 To make the computation easier, we assume a pyramidal shape for  the jet 
 with a square basis of side $2 r_0l$ and with a lateral surface facing the line of sight. 
In that way the integration path inside the jet does not depend on other parameters related to the curved jet surface.
The perpendicular direction to this surface makes  an angle $[90^o-(\eta - \xi)]$ with the line of sight for the case of the approaching jet and  $[90^o-(\eta + \xi)]$ for the receding jet  (see Fig.~A1-top in the Appendix).

With these assumptions   the  approaching jet emission can be written as
\beq
S_{\rm a}=\int^L_1   r_0 x_0 I_{\nu_{\rm a}}(\eta, l)  {\sin (\eta- \xi) \over D^2} l {\rm d}l 
\eneq
where $D$ is the distance of \lsi, and the integral is
from $l=1$ to $l=L$,  with  $x_0L$ being the jet length 
and $L$ operationally determined 
as the limiting value  above which 
there are no more contributions  to the jet flux. 
For the receding jet, we have the same expression with a change of sign in the angular factor that takes
the projection of the emitting surface in the direction perpendicular to the line of sight into account.
Clearly the  difference between the emission of the approaching and  of the receding jet
becomes negligible for cases where $\xi<<\eta$.

The intensity $I_{\nu}$,  emerging from the jet surface, i.e.,  at $\tau =0$, in the direction  $\eta$, is
  deduced from the radiative transport equation when considering that the plasma inside the jet is stratified along the jet axis direction, i.e., along a direction that  makes an angle $\eta$ with respect to the line of sight 
\beq
I_{\nu }(\eta, l)=  \int _{0}^{\tau_{\rm end}(l)} {J_{\nu} \over \chi_{\nu}}e^{- \tau' / \cos~\eta} ~{\rm d}  \large  \left [{\tau' \over \cos~\eta} \large \right ]
,\eneq
where $\tau$ is the optical depth  defined as
\beq
\tau (l)= -  \int^{\infty}_l  \chi _{\nu} ~ {\rm d}x =  \tau_0 l^{1-a_3-(p+2)a_2/2}.
\eneq
 The upper integration limit  $\tau_{\rm end}(l)$  
  represents the jet   optical depth  at the  specific value of $l$
and takes  into account that,   at each distance, $x_0 l$, from the beginning of the jet,
 the integration path inside the jet involves 
different regions of the jet itself (see Appendix).  The optical depth inside the jet is also a function
of the angle $\eta$ (see  expressions (A.5) and (A.6)) inducing  higher values of the optical depth 
when the jet is pointing toward the observer, i.e. for low $\eta$ values (but in the limit
 $\eta \ge  \xi$). The consequences of this dependence  will be analysed further  in a subsequent paper.

The exponent for $l$ in expression (8), when taking Kaiser's Table 1 into account for all possible $a_i$ values,
is always negative, so  the optical depth is at its maximum at the beginning of the jet:
\beq
\tau_0= \tau(l=1)={ \chi_0 x_0 \over -[1-a_3-(p+2)a_2/2]} \nu^{-(p+4)/2}.
\eneq

Emission and absorption  quantities in the above expressions are given, as in \cite{longair}  and following the  \cite{kaiser06} 
parametrization, by
\beq
\unit[ J_{\nu}= J_0  \nu^{-(p-1)/2} l^{-a_3-(p+1)a_2/2}]~~~~{{W \over {\rm m}^{3}{\rm Hz}}}
\eneq
\beq
\unit[ \chi_{\nu}= \chi_0  \nu^{-(p+4)/2} l^{-a_3-(p+2)a_2/2 }]~~~~ {{\rm m^{-1}}}
,\eneq
where 
\beq
J_0=2.3~10^{-25}(1.3~ 10^{37})^{(p-1)/2}a(p)B_0^{(p+1)/2}\kappa_0
\eneq
and
\beq
\chi_0=3.4~10^{-9}(3.5~10^{18})^p b(p)B_0^{(p+2)/2}\kappa_0.
\eneq
with the constants $a(p)$ and $b(p)$ 
as in \cite{longair}
and $\kappa_0$ as  determined in the next section. 

\begin{figure*}[ht]
\includegraphics[width=13.cm, height=11.cm,angle=0]{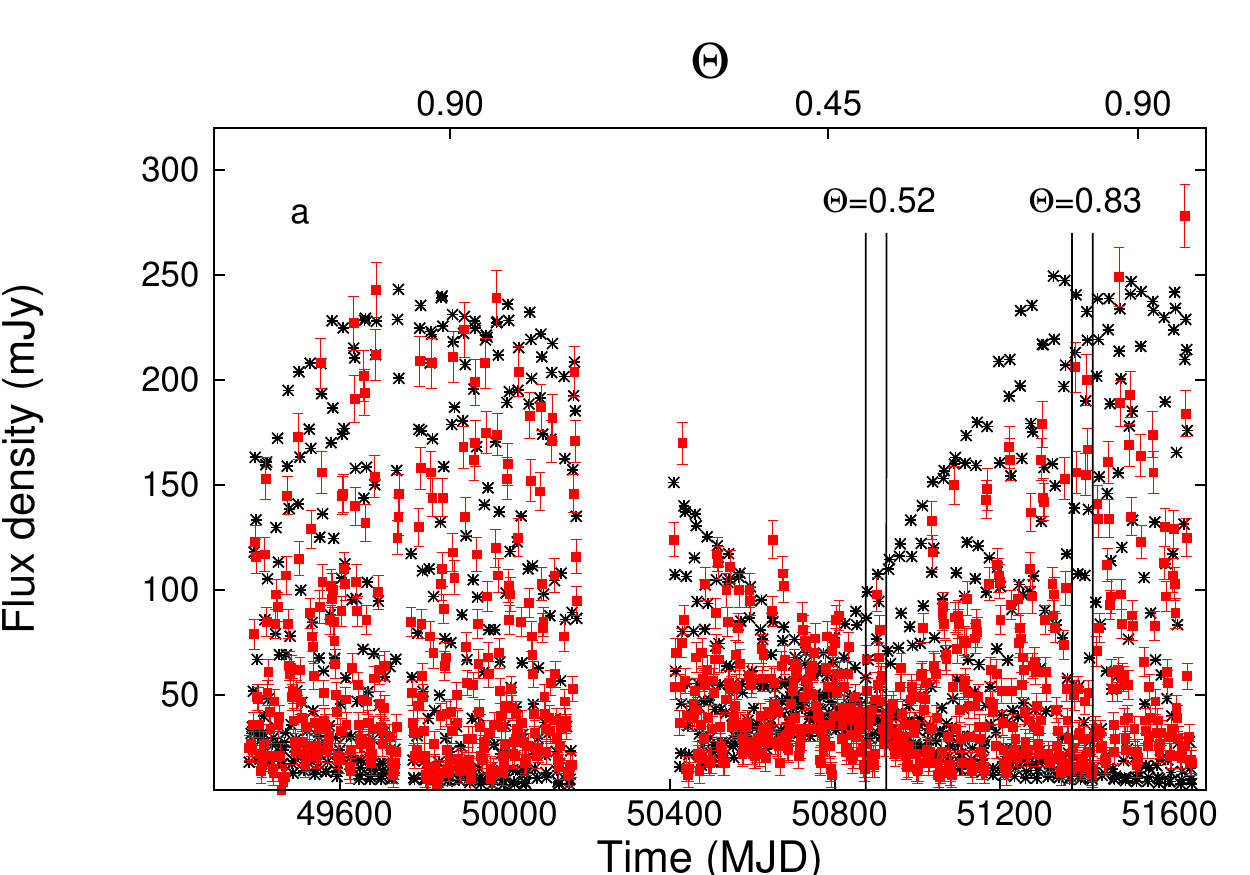}\\
\centering
\includegraphics[width=7.cm,angle=0]{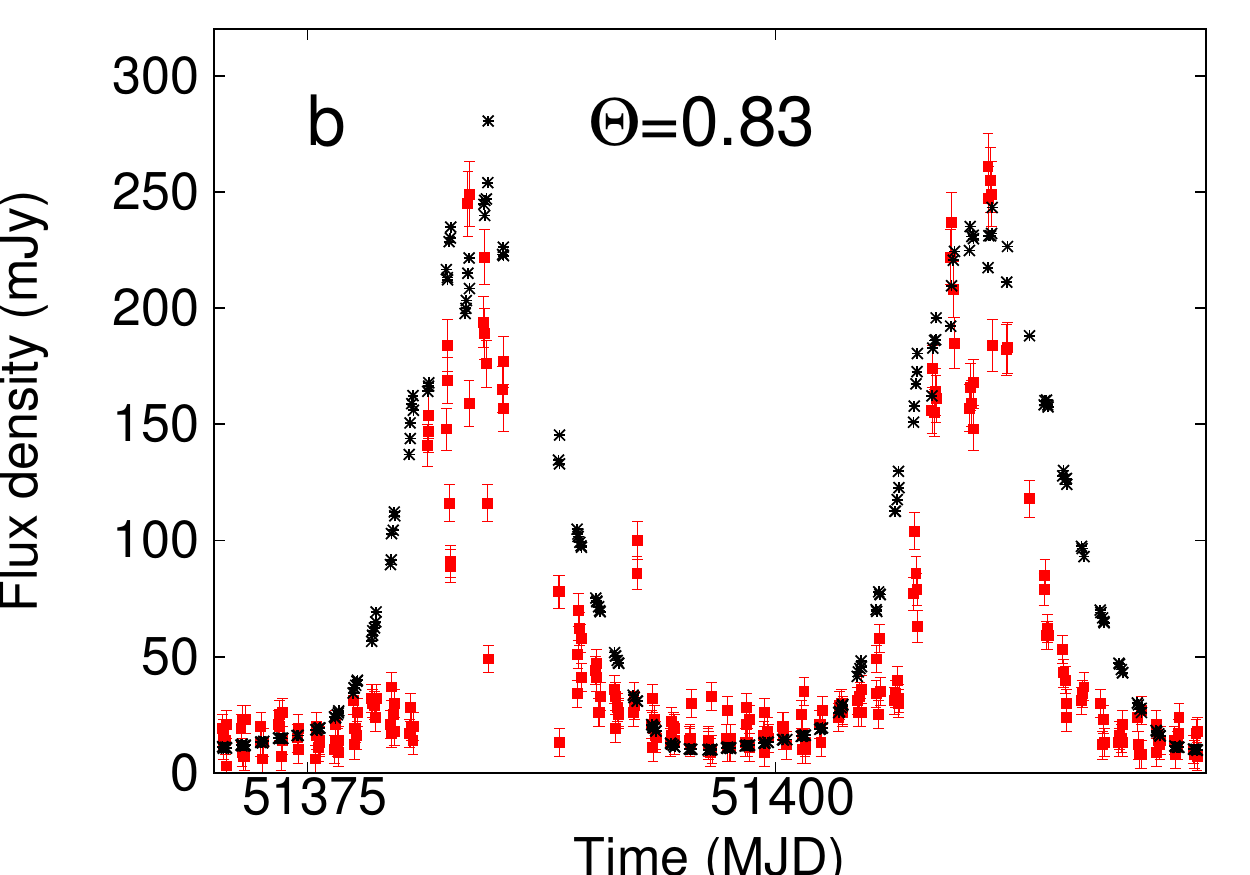}
\includegraphics[width=7.cm,angle=0]{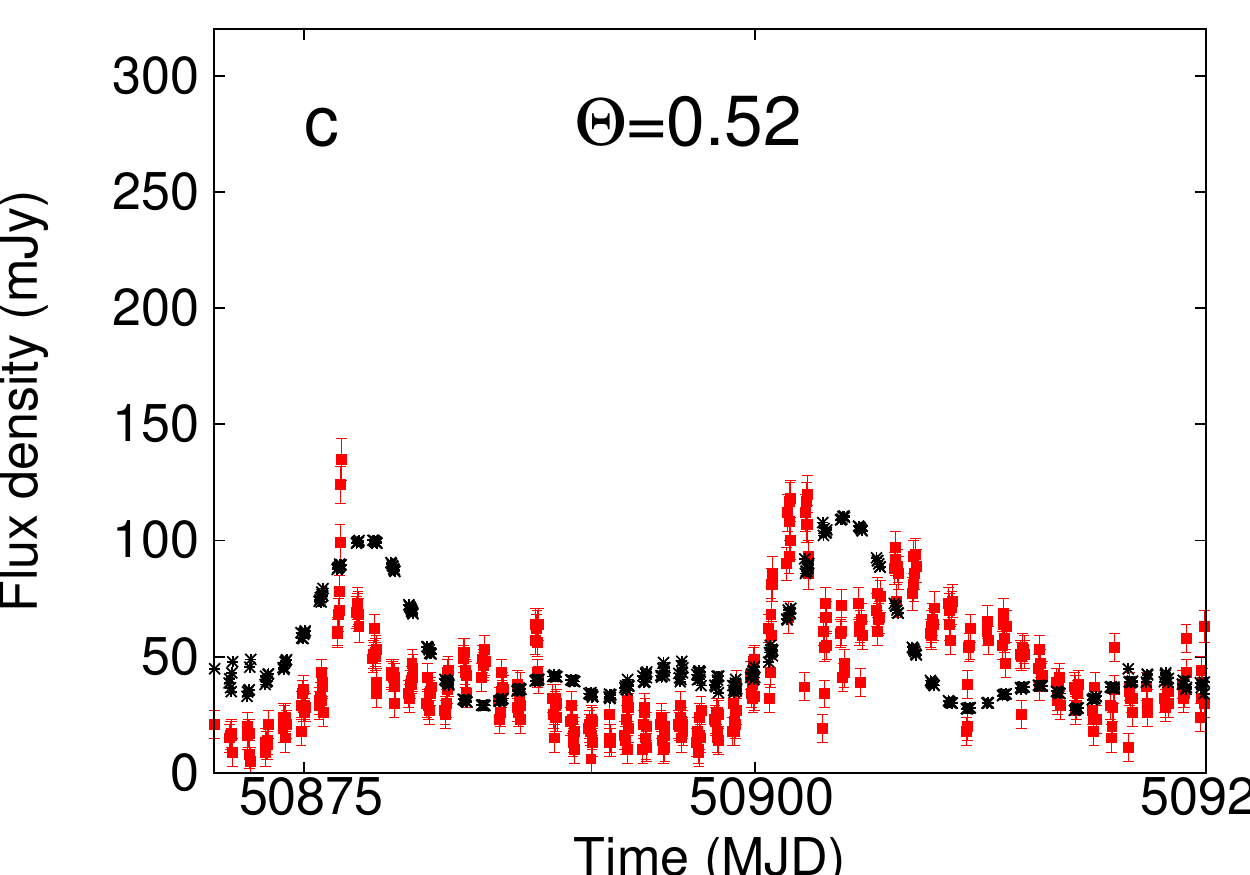}\\
\caption{Long-term modulation in \lsp.  8.3~GHz GBI observations (red/squares) and model (black/stars) data. 
a): Model and data  vs time (bottom x-axis) and vs the long-term modulation phase $\Theta$ (top x-axis).
The long-term modulation phase $\Theta$  is the fractional part of $(t-t_0)/1667$    
with     $t_0$=JD2443366.775  \citep{gregory02}. The GBI data are averaged over 2 days. 
The four bars define the two time intervals zoomed in Figs. 2\,b and 2\,c.
 b) and c): Zoom of two intervals of Fig.\,2\,a.
  This (Fig. 2)  and all other figures (3, 4, 5, 6, and 7)  have been obtained for $B \equiv B_{\parallel}$,
$\zeta=46$, $\psi=40$, $\Delta=-922.5$, $\beta=0.28$, $A=3.84$, $\Phi_r=0.6$, $L=14$, and $\xi=1.5$ 
 (Sect. 3). 
}
\label{mod}
\end{figure*}
\subsection{Periodical increase  in relativistic particles  along the orbit}
 To take into account the observed orbital modulation in the radio flux
 we make the working hypothesis that the relativistic electron  input in the jet changes with time. 
This hypothesis is plausible since  as discussed below,  
the accretion is variable along an eccentric  orbit. 
Our type of representation 
 therefore includes the classical hypothesis that the presence of relativistic
electrons        
 in the jet is related to  matter accretion on the compact object 
 but makes no hypothesis for how they have been accelerated. 

For  \lsp, with $e=0.72 \pm 0.15$ \citep{casares05},  the \citet{bondihoyle44}  accretion  theory in an eccentric orbit
predicts two maxima  \citep{taylor92, martiparedes95, boschramon06, romero07}.
 One maximum is around periastron at $\Phi_{\rm periastron}=0.23 \pm 0.02$
\citep{casares05},
with the orbital phase $\Phi$  defined as the fractional part of    
 ${(t-t_0)\over P_1}$. 
However, near periastron the ejected relativistic electrons  suffer severe inverse
Compton losses because of the proximity to the Be star;  a  high energy 
outburst is expected but no or  negligible radio emission \citep{boschramon06}. 
 For the second accretion peak, the
compact object star is much farther from the Be star, and inverse Compton losses
are lower. The relativistic electrons can propagate out of the orbital plane, and we observe a radio outburst 
\citep{boschramon06}.

In modelling the radio emission we assume a periodical
function for relativistic electron density:
\beq 
\kappa_0=A~{\it f}(\Phi-\Phi_{\rm r})
\eneq
where ${\it f}(\Phi-\Phi_{\rm r})$ is equal to 1 for $\Phi=\Phi_{\rm r}$
and  expresses (times $A$) the periodical ($P_1$) 
increase in relativistic electrons due to Bondi's accretion in an eccentric orbit. 
   In this first stage (focussed on Doppler boosting   effects
and on data 
periodicity reproduction), we do not include
       relativistic electron energetic losses but instead allow  a different 
  slope for the onset and for  the decay so as  to take into account that electrons 
take some time to lose their energy.   
 The top of Fig.~\ref{shift} shows 
 the resulting $\kappa_0$ function  for the fit solution  illustrated in Figs.~2-7.

\begin{figure*}[ht]
\centering
\includegraphics[width=8.0cm,angle=0]{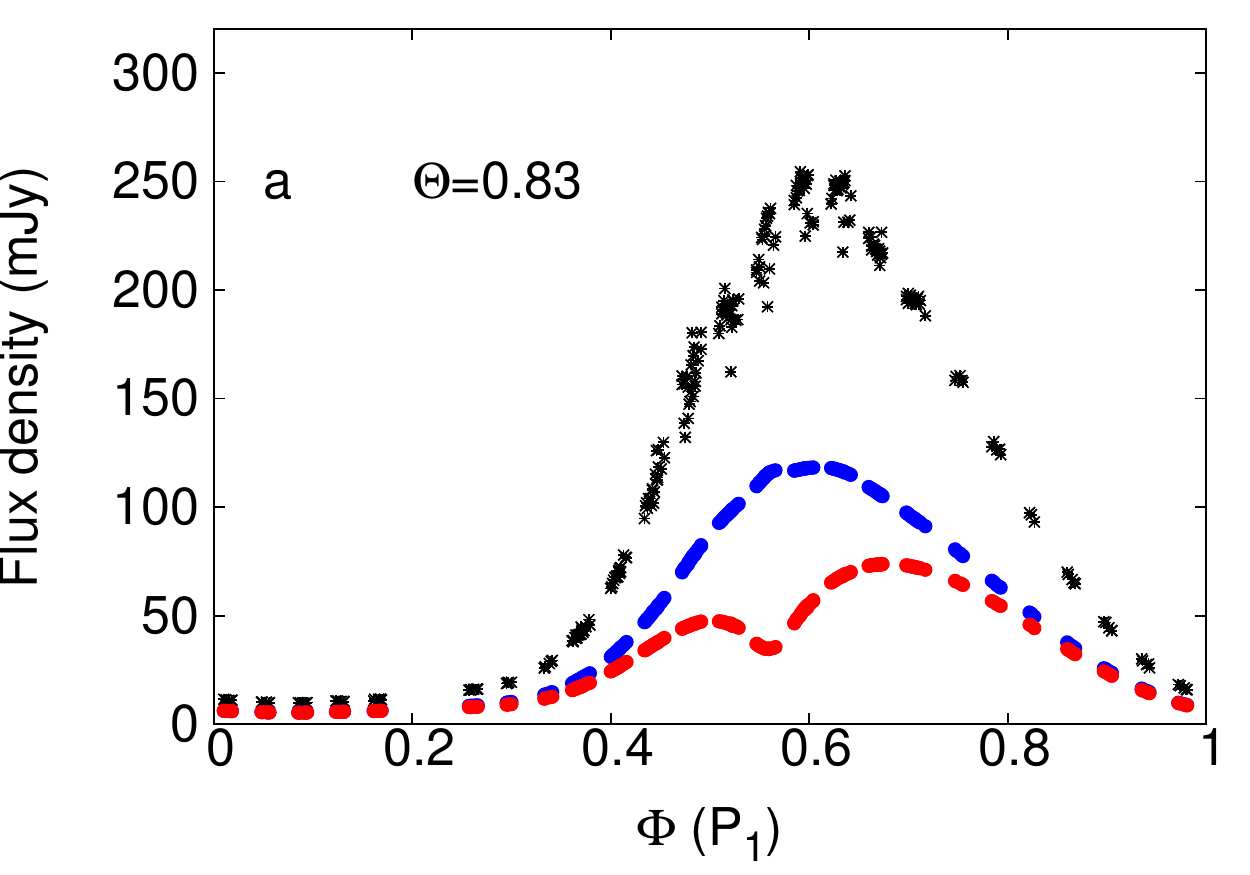}
\includegraphics[width=8.0cm,angle=0]{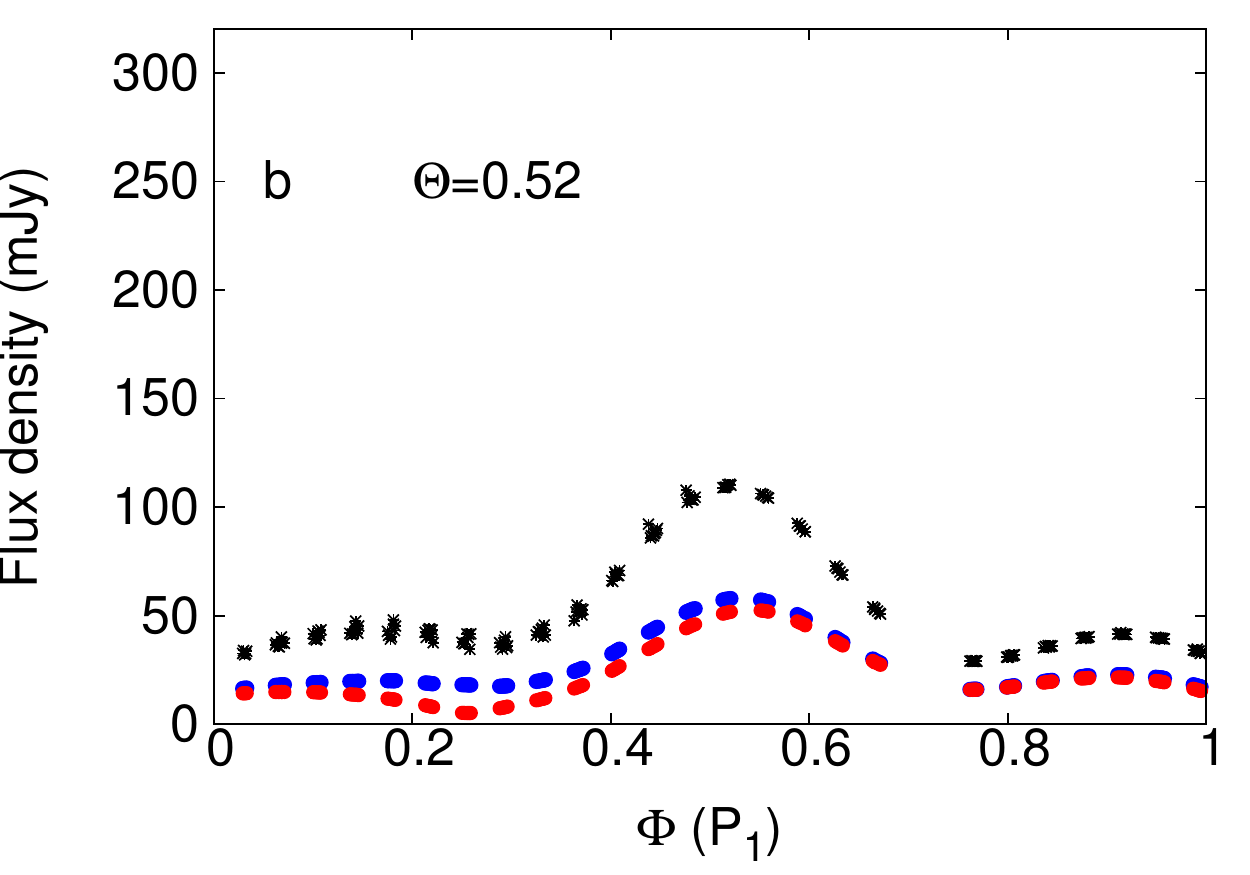}\\
\includegraphics[width=8.0cm,angle=0]{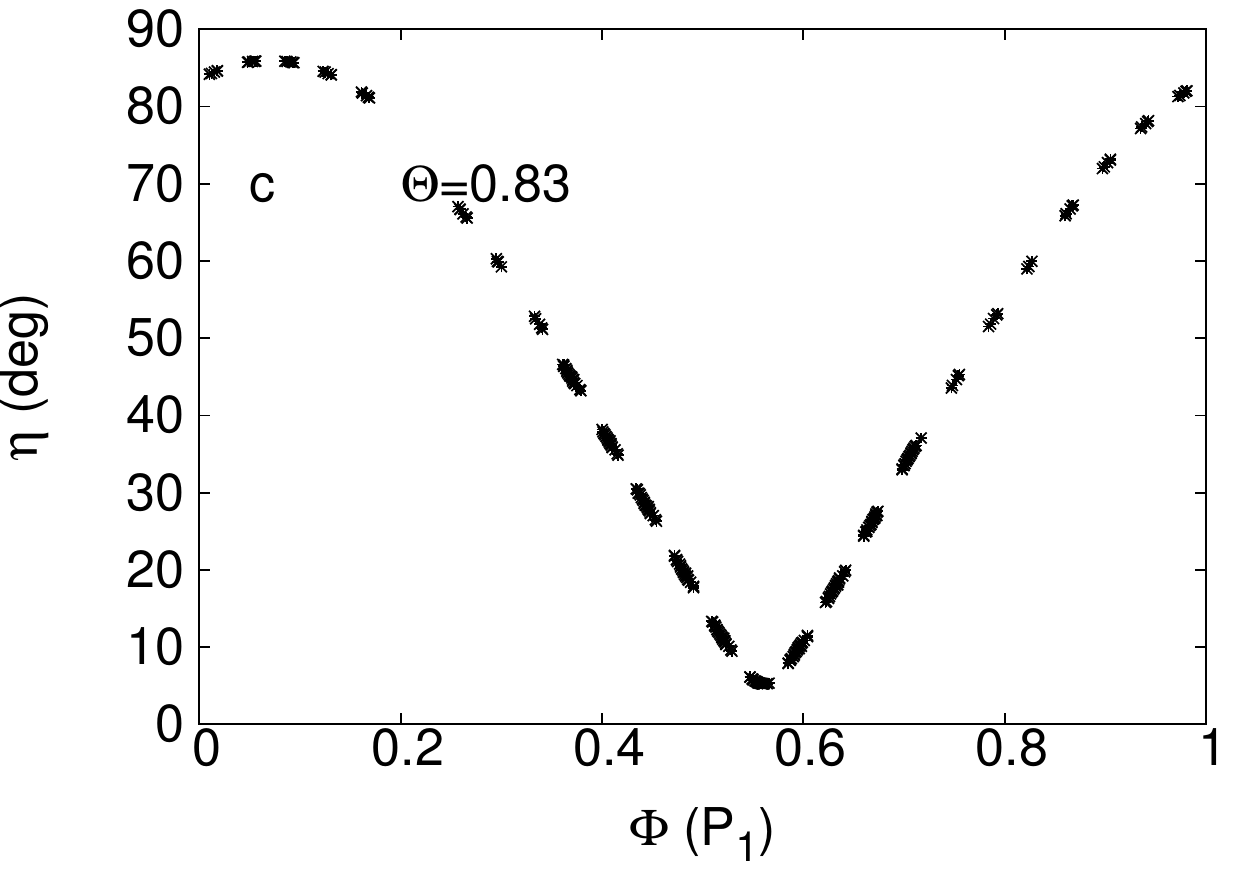}
\includegraphics[width=8.0cm,angle=0]{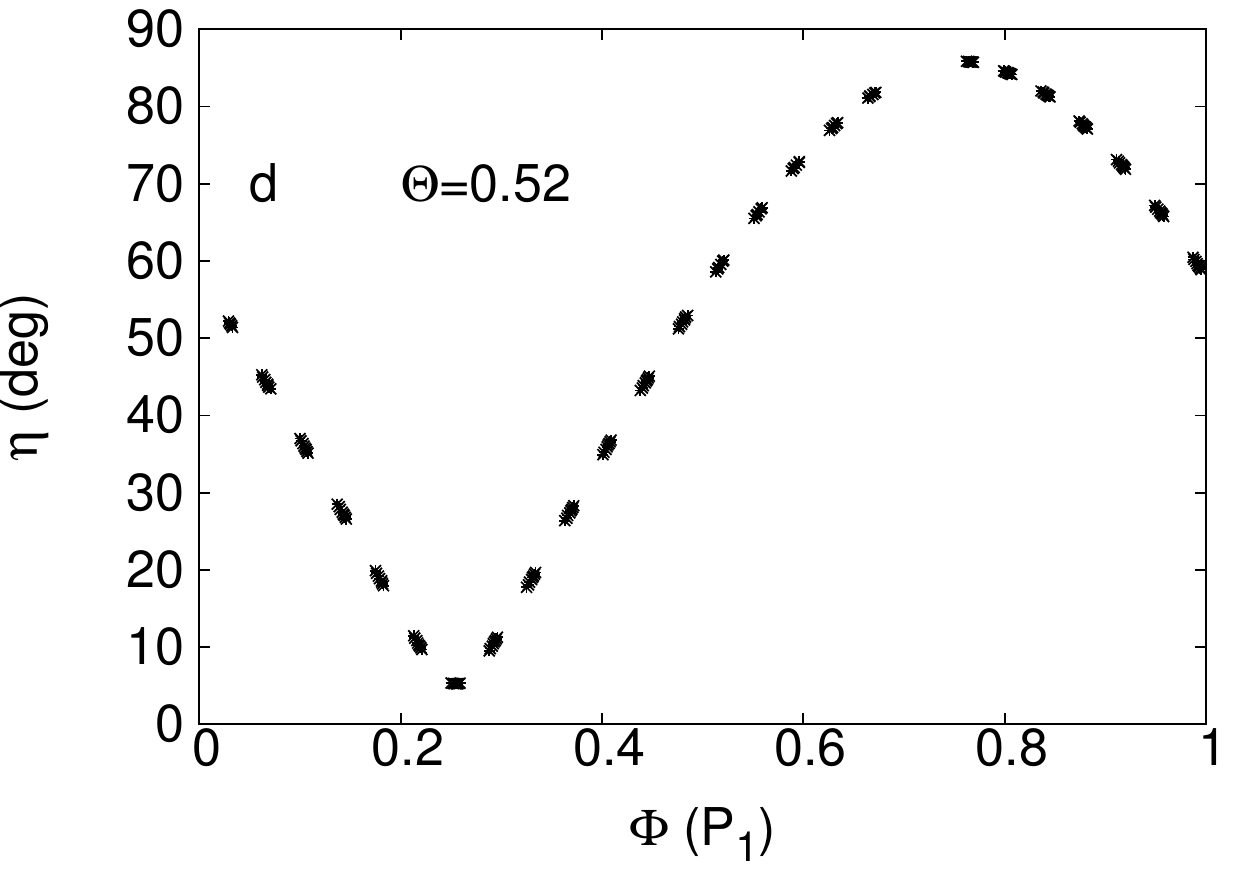}\\
\includegraphics[width=8.0cm,angle=0]{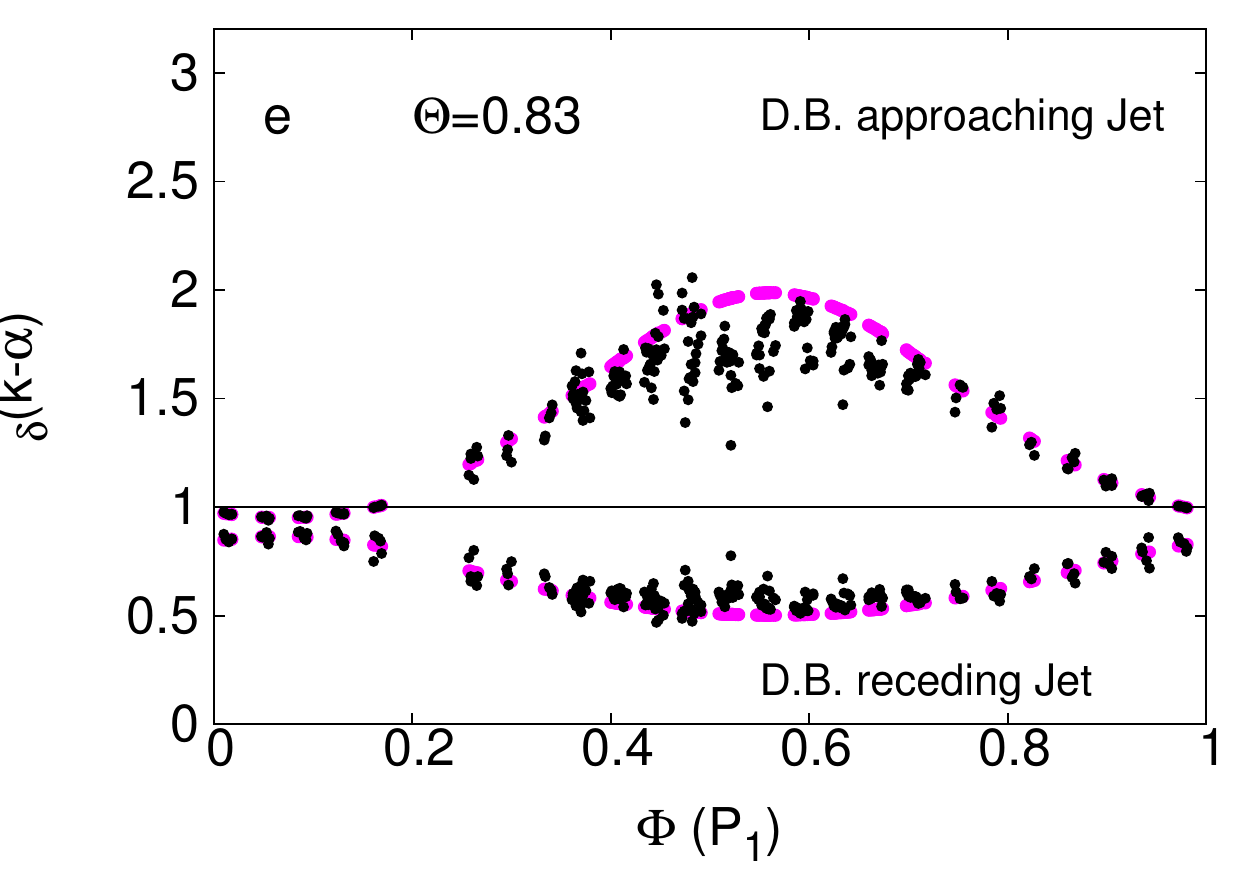}
\includegraphics[width=8.0cm,angle=0]{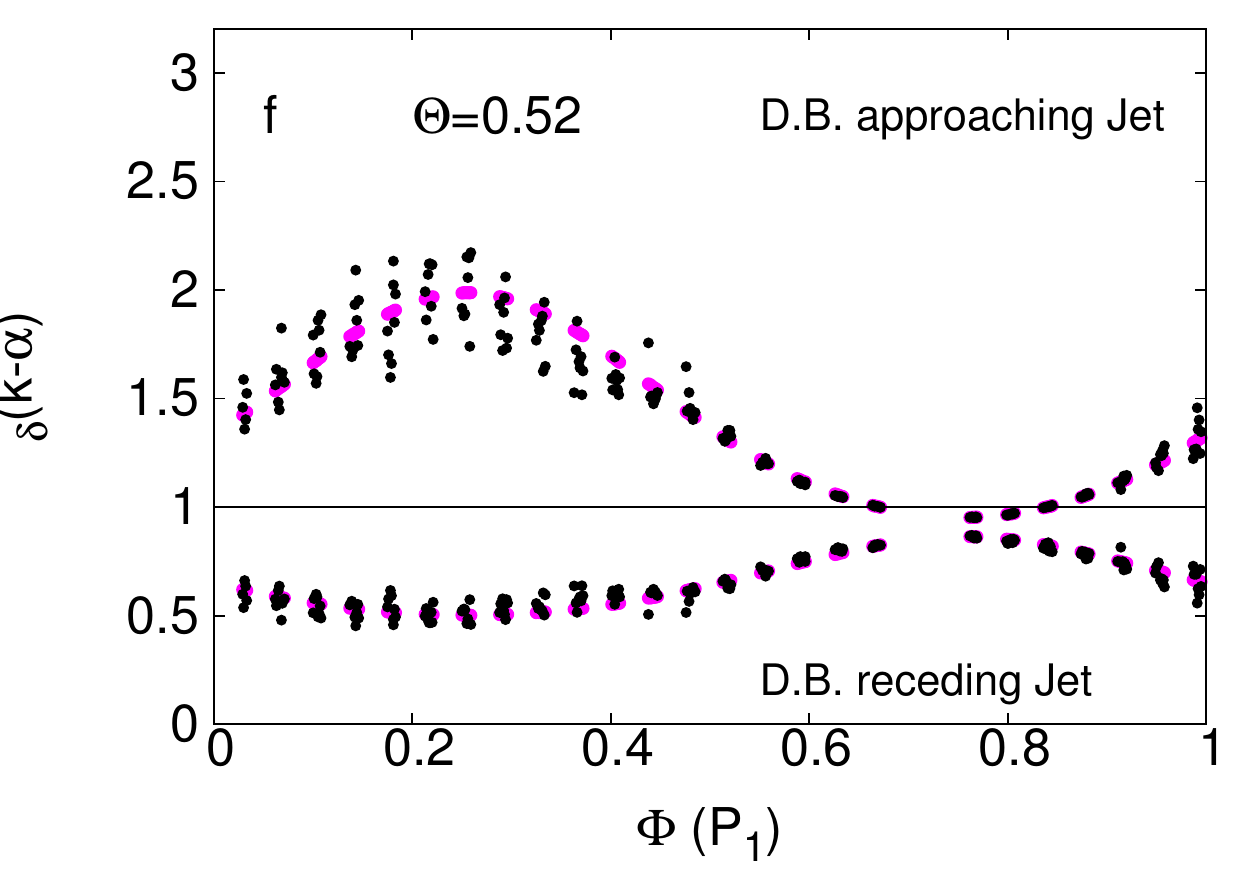}\\
\caption{Doppler boosting and intrinsic emission. a-b) $S_{\rm model}$ (black stars), $S_a$ (blue line) and $S_r$ 
(red line) of Eq. 1 vs orbital phase $\Phi$, for long-term modulation phases $\Theta=0.83$ and $\Theta=0.52$.
c-d) The angle $\eta$ between the  jet axis and the line of sight vs orbital phase $\Phi$.
e-f)  Doppler boosting factors computed with observed spectral index values  $\alpha$. The violet curves correspond
to Doppler boosting factors computed for $\alpha=-0.4$ (see Sec.4.1).}
\label{dopp}
\end{figure*}
 
\section{Model fitting}

Table 1 reports the  parameters, described in the previous section, 
needed to calculate,  by Eq. (1),   the flux density of our model, $S_{\rm model}$.
The values for the parameters  were found by minimizing 
the  $\chi^2$  between $S_{\rm model}$, computed at a given $t_i$,  and 
the simultaneous observed flux density with the Green Bank interferometer at
8 GHz.  
That is, we found  those values of
the parameters that give the lowest value of $\chi^2$:
\beq
\chi^2 =\sum_i {(S_{\rm GBI}(t_i) -S_{\rm model}(t_i,parameters))^2\over \sigma_i^2}\eneq
where $\sigma$ is the error of GBI data at 8 GHz.

We used the  function minimization software MINUIT of the CERN Program Library.
In particular, we first set an initial model using the program SCAN, 
then we optimize the model with the  program  SIMPLEX  to perform    
the  minimization using the simplex method of \citet{neldermead65}.
        The best solution is for $B \equiv B_{\parallel}$, 
$\zeta=46$, $\psi=40$, $\Delta=-922.5$, $\beta=0.28$, $A=3.84$, 
$\Phi_r=0.6$, $L=14$, and $\xi=1.5$.  
All figures (Figs. 2, 3, 4, 5, 6, and 7) have then been made with this model. 
However, the non-linearity of the problem makes the solution depending 
on the initial model,
and  we also found other solutions with only slight worse  fits.
This uncertainty is considered in the parameter errors (shown in Table 1) that  give 
the dispersion with respect to the mean value 
found for each parameter in the different solutions. 
The  solutions for 
the different magnetic configurations, i.e. cases with 
$B\equiv B_{\parallel}$  or  $B\equiv B_{\perp}$ are  within the given errors.
The solutions in Table 1 therefore apply to both cases of $B$. 

\begin{table*}
\center
\begin{tabular}{lcccccccc} 
$\zeta$& $\psi$& $\Delta$&$\beta$&A&$\Phi_r$&L&$\xi$\\
($^{\circ}$)&($^{\circ}$)& (d)&&&&&($^{\circ}$)\\
\hline
47$\pm$17&29$\pm$12&-922$\pm$1&0.45$\pm$0.18&3.0$\pm$1.5&0.58$\pm$0.02&13$\pm$2&2.8$\pm$1.5\\
\hline
\end{tabular}
\caption{Parameters of the precessing conical jet derived by the fitting procedure of Sect. 3: 
$\zeta$ is the  angle between jet precession axis and the line of sight, 
$\psi$  the  opening angle of the precession cone,
$\Delta$  the  offset  to synchronize $P_2$ to $P_1$,
$\beta$ the  jet velocity parameter, $A$ and   $\Phi_r$  are  parameters 
of the periodical function of the relativistic electron of Eq. 14, 
$L$ is the jet length, and  $\xi$ the jet opening angle.}
\end{table*}

\section{Results and discussion}

We compare here model data and observations, both  given as a function of time and  of the phases, as the  orbital phase $\Phi$ and
 the  long-term modulation phase $\Theta$.
Model data have been generated with the parameter values as specified in Fig. \ref{mod} caption.
 
\subsection{Long-term modulation and Doppler boosting effects}
Figure \ref{mod}\,a  shows  GBI data (averaged over 2 days) and model data, i.e. $S_{\rm model}$,  vs time.
The model  reproduces  the data at this "large scale". 
Now we analyse  the plots in detail. In Fig.~\ref{mod}\,a, two bars are given around 
the minimum of the long-term modulation around  50900~MJD
and two other bars at a higher flux density phase of \lsi around  51375 MJD. 
In terms of the long-term modulation phase $\Theta$, 
the two time intervals correspond to $\Theta=0.52$ and  $\Theta=0.83,$ respectively.

One can see the zoom of the outbursts at  $\Theta=0.83$ 
 in Fig.~\ref{mod}\,b and how the model  also reproduces  on a  "fine" scale the periodical large outburst occurring at that
$\Theta$ phase.
The  model data   are folded as a function of the orbital phase $\Phi$ in  Fig.~\ref{dopp}\,a. Along  with 
$S_{\rm model}$ of Eq. 1, we also plot  the intrinsic  emission of the approaching jet $S_a$ (blue points)
 and that of the receding jet $S_r$ (red points). 
We  note that the intrinsic emission of the two jets is different.
This difference between the approaching and the receding jet,  which is due to 
 optical depth effects and is important  for low values of $\eta$, 
 will be analysed in detail  in a subsequent paper.
We now discuss the effects of  Doppler boosting.
Figures~\ref{dopp}\,e  and \ref{dopp}\,f   show (black points)  the    
Doppler factors $\delta ^{{\rm k} -\alpha}$ for  $k=2$ and   $\alpha(t)$   
calculated from  GBI observations  at 2.2 GHz and 8.3 GHz.
As a comparison,  the violet curve in the figures corresponds to  Doppler factors with 
$\alpha=(1-p)/2$ and  $p$ equal to the value  
$p=1.8$  used in our model for the electron energy distribution (see Sect. 2.1).  
As  can be seen  in Fig.~\ref{dopp}\,c,
the minimum  $\eta$ value, which determines the maximum Doppler boosting,
  occurs at $\Phi= 0.56$, which is rather close to the orbital phase $\Phi= 0.6$ 
 of the peak  of the intrinsic emission  $S_a$  (Fig.~\ref{dopp}\,a, blue points).
This coincidence of strong intrinsic emission and maximum Doppler boosting   
gives rise to the large outbursts observed at this
$\Theta$ phase in \lsp.   

We likewise examine the relationship  between Doppler effects and the minimum of 
the long-term modulation.
One can see the zoom  at about 50900~MJD in Fig.~\ref{mod}\,d. 
In the two consecutive outbursts,
the model  reproduces the periodicity and the lower  amplitude of the outbursts.
The model data  folded  
in function of the orbital phase are given  in  Fig.~\ref{dopp}\,b. 
As shown in Fig.~\ref{dopp}\,d,
the angle $\eta$ between the jet and the line of sight has its minimum value at about $\Phi$=0.25.
This implies that at this $\Theta$ phase  the maximum Doppler boosting  (Fig.~\ref{dopp}~f) occurs in an unfavorable situation, i.e., 
when there is little emission to amplify (low values of  $S_a$  and $S_b$ in Fig.~\ref{dopp}\,b). 

\begin{figure}[ht]
\includegraphics[width=9.cm,angle=0]{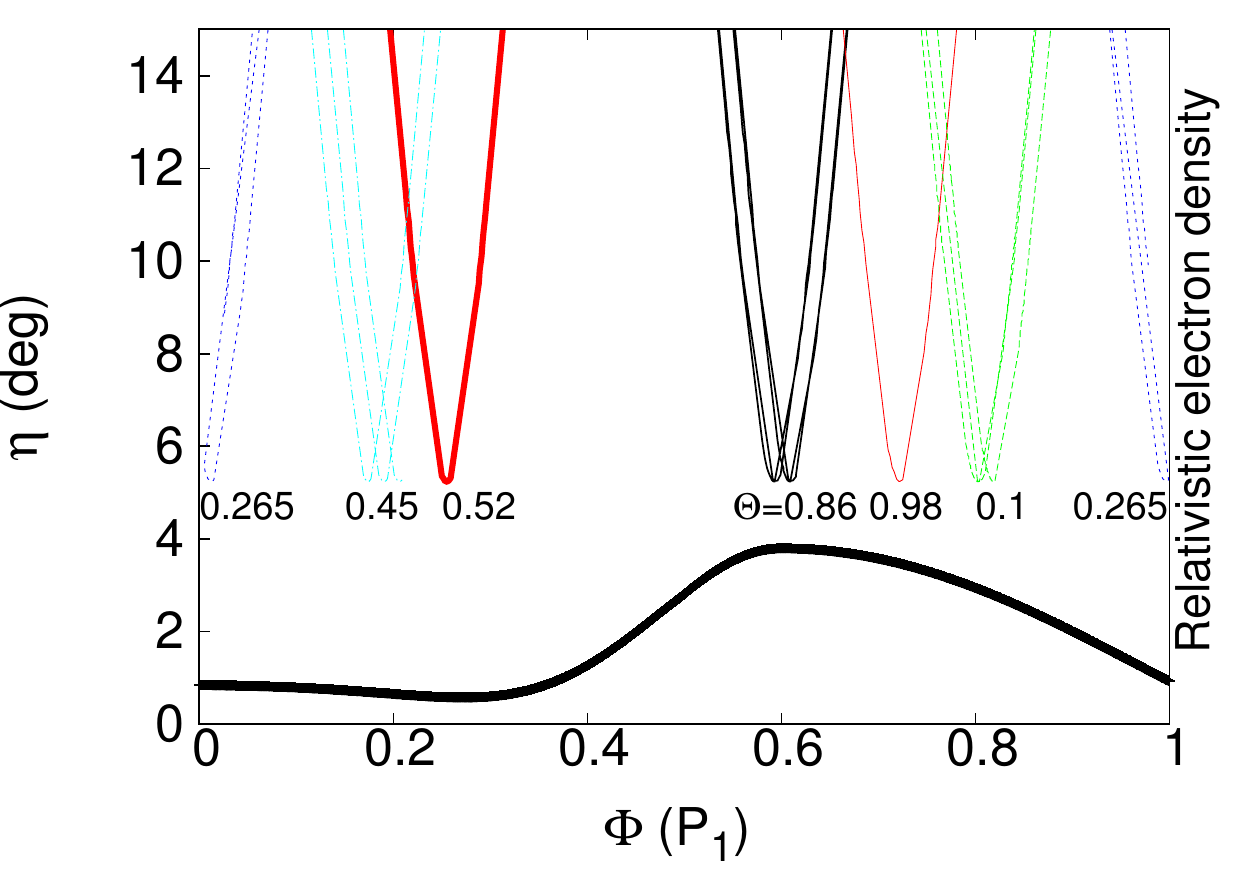}\\
\includegraphics[width=9.cm,angle=0]{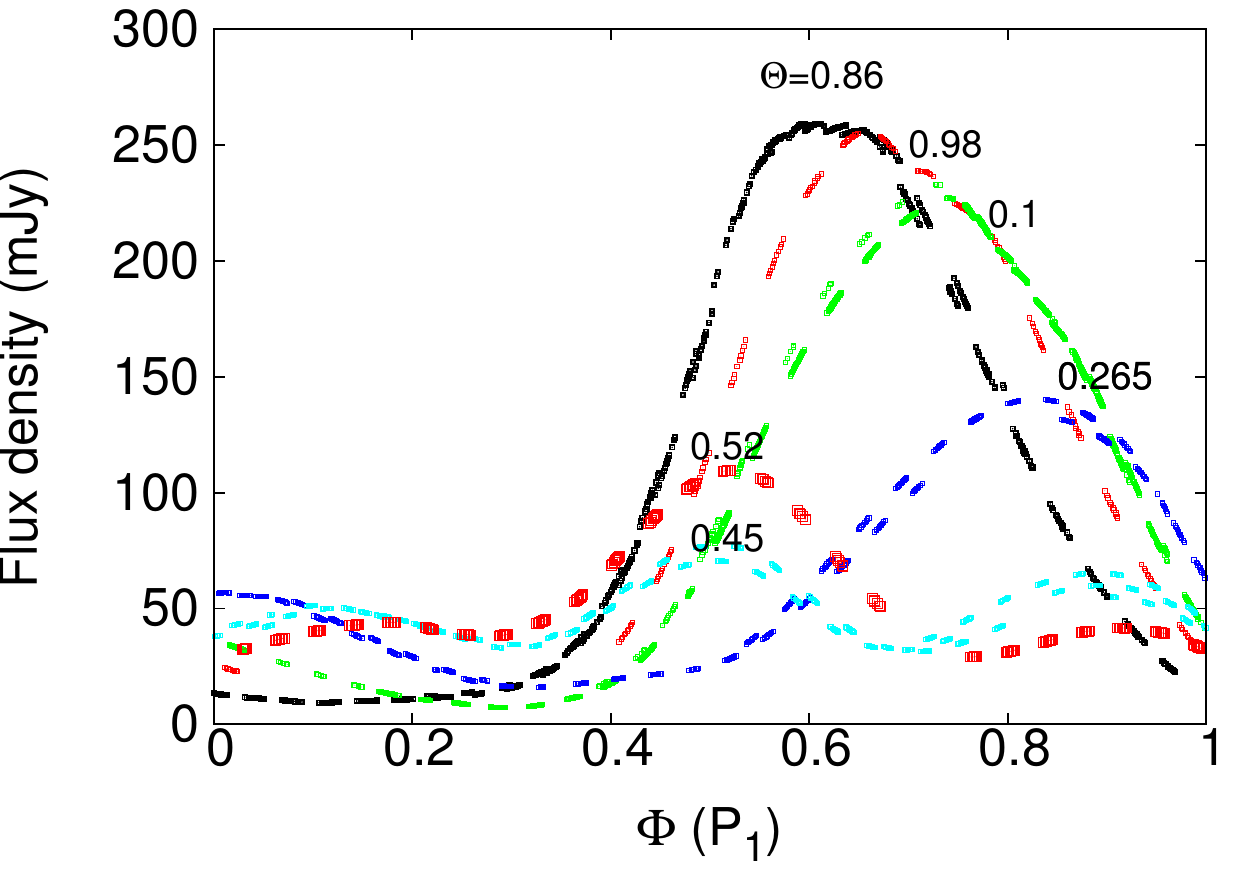}\\
\caption{
Top: The angle $\eta$ between jet axis and line of sight at different
phases of the long-term modulation ($\Theta$) vs orbital phase ($\Phi$). 
One sees as the minimum of
 $\eta$, which corresponds to maximum Doppler boosting, moves at different $\Phi$s 
for the different $\Theta$ phases. The black thick curve  is (on an arbitrary  scale) the adopted relativistic 
electron density function of $P_1$ of Sect. 2.4. 
Bottom: 
Amplitude modulation and orbital shift of the outbursts: 
outbursts at different $\Theta$s vs $\Phi$ (with the same colour code as the above panel).
The outburst decreases its amplitude and shifts
in orbital phase. 
} 
\label{shift}
\end{figure}

\begin{figure}[ht]
\includegraphics[width=8.cm,angle=0]{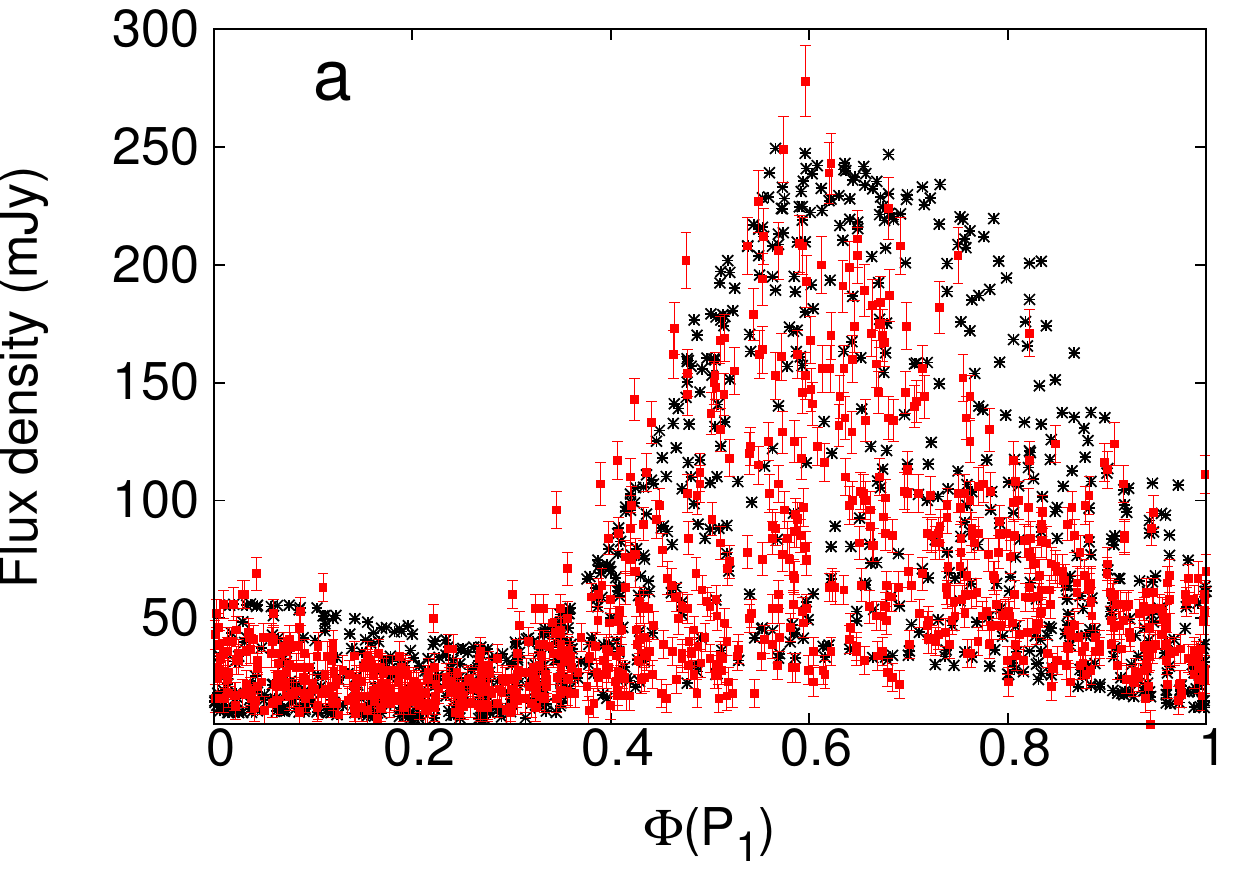}\\
\includegraphics[width=8.cm,angle=0]{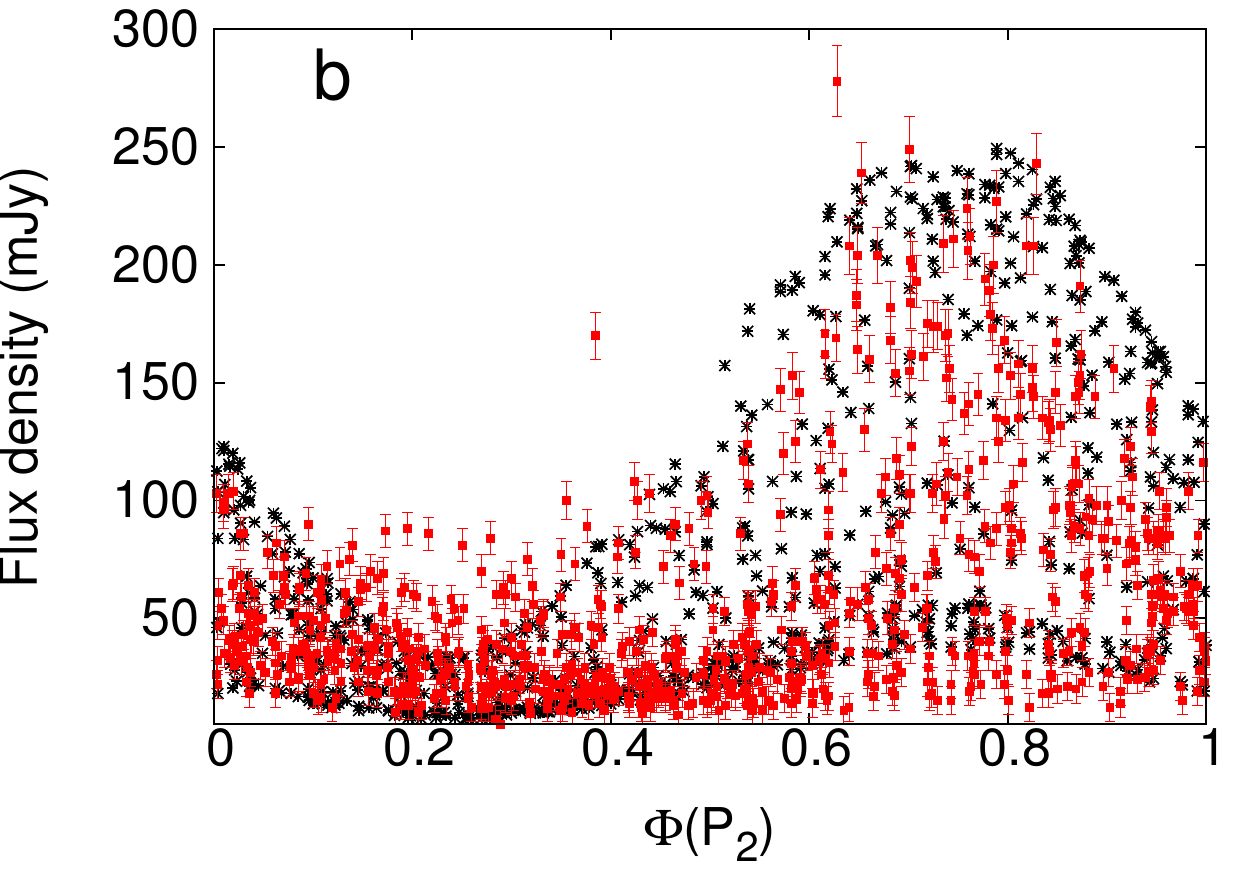}\\
\includegraphics[width=8.cm,angle=0]{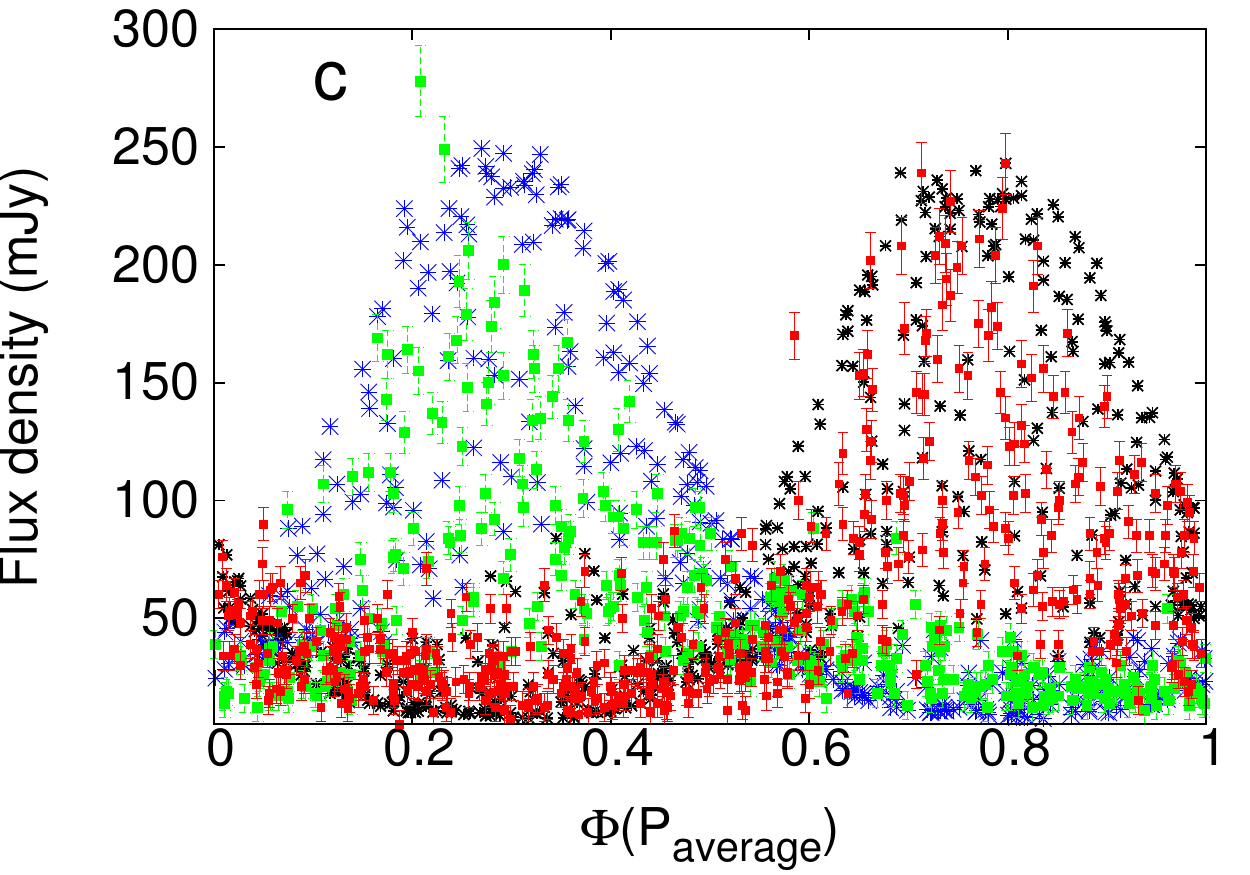}\\
\caption{ 8.3~GHz GBI observations (red squares) and model (black stars) data. 
All phases, $\Phi$, refer to  $(t-t_0)$, 
with     $t_0$=JD2443366.775  \citep{gregory02}.
a): vs $\Phi$ ($P_1$)  (orbit        $P_1=1/\nu_1$=26.49 days); 
b): vs $\Phi$ ($P_2$) (precession    $P_2=1/\nu_2$=26.92 days;  
c): 
 8.3~GHz GBI observations 
before (red points) and after (green points) 50841 MJD vs $\Phi$ ($P_{\rm average}$)   ($P_{\rm average}=2/(\nu_1+\nu_2)$=26.70 days).
The points cluster separately 
with a shift of 0.5 in phase. The model reveals  the same double clusterings: Black
points refer to model data before 50841 MJD and blue points to model data after 50841 MJD.  
}. 
\label{beat}
\end{figure}
\subsection{Amplitude modulation and orbital shift of the outburst occurrence}
During the maximum of the long-term modulation, the peak of the outburst 
occurs at orbital phase $\Phi \sim 0.6$ \citep{paredes90}.
Afterwards, the orbital phase of the peak of the outburst changes, a phenomenon analysed  in the past 
by \citet[with references there]{paredes90} in terms of orbital phase shift, and 
by \citet{gregory99} in terms of timing residuals. 
In this section we show how our model of a precessing ($P_2$) jet with periodically ($P_1$)
varying ejection reproduces  i) the observed gradual shift of outburst occurrence
from $\Phi=0.6$  
  to  longer orbital phases ($\Phi\sim 0.9$) towards  the minimum, 
ii) the so-called ``quiet phase emission with no distictive maximum" discussed in \citet{paredes90},
and iii) the observed re-appearing of outbursts having a distinctive maximum, even if still low,
at orbital phase $\Phi \sim 0.5$.

Figures~\ref{dopp}\,c and~\ref{dopp}\,d discussed in the previous section show the variations of angle $\eta$
formed by the jet with the line of sight, vs  orbital phase 
for the two extreme situations: $\Theta=0.83$ and 
$\Theta=0.52$,
which is close to  the maximum and the minimum of the long-term modulation.
 We now  consider
the curves  of  $\eta$ for several different $\Theta$ values (Fig. \ref{shift}\,top) and the related  outbursts  
(Fig. \ref{shift}\,bottom). 

At $\Theta$=0.86, the minimum of $\eta$ that corresponds to the  maximum Doppler boosting
occurs at $\Phi$=0.6. 
In the same Fig.~\ref{shift}\,(top)  the adopted relativistic 
electron density is drawn, as a function of $P_1$, 
with its maximum at $\Phi$=$\Phi_r$=0.6. 
Clearly at $\Theta$=0.86 the maximum Doppler boosting and the maximum ejection occur at the same orbital phase;
that is, the electrons are ejected at the minimum $\eta$ angle with respect to the line of sight. 
After 200 days, at $\Theta$=0.98 the maximum Doppler boosting  occurs at $\Phi$=0.7 (red curve 
at  $\Theta$=0.98 in Fig.~\ref{shift}\,top), and  it matches the decay phase of the ejection. 
The result is the red dotted curve of Fig.~\ref{shift}\,(bottom): an outburst slightly lower than 
that at $\Theta$=0.86
and shifted at $\Phi$=0.7.
At $\Theta$=0.1 we can appreciate the evolution of the same phenomenon:
the maximum Doppler boosting (MDB) (green curve in Fig.~4\,top) is further shifted in phase and this causes 
a shifted and lower outburst (green curve in Fig.~4\,bottom).
In the interval starting after  $\Theta$=0.265 (with the peak of the outburst at $\Phi \simeq 0.85$)
 until   $\Theta\sim$0.45
, the outburst  loses its typical shape, 
which is now a rather broad low outburst, and it becomes very difficult to define a peak 
(see the curve at $\Theta=0.45$ in Fig.~\ref{shift}\,bottom).  
This is the so-called ``quiet phase emission, with no distinctive maximum" 
noticed by  \citet{paredes90}
during the minimum of the long-term modulation.
Finally, the outburst starts to resume its  shape at  $\Theta$=0.52.
As one can see in Fig. 4,
there is  some boosting  at  the onset of the new ejection peaking at $\Phi=0.6$.
The result is an outburst with peak  at $\Phi\simeq 0.5$.
Our model, therefore, is able to reproduce the  orbital phase shift of Fig.~5 of 
\citet{paredes90}.

\subsection{Orbital and outburst periodicities}
Figure \,2\,b,  where
the fits of  two consecutive large outbursts are shown,
has already proved the ability of the model
also  reproducing the shorter orbital periodicity, $P_1=26.49$\,d.
Here the comparison is extended to  the whole data set,
by folding, as shown  in Fig.  \ref{beat}\,a,
all 6.7 yr  of GBI data   and  model data with  $P_1$.
However,  $P_1$ refers to  the orbital periodical ejection of relativistic electrons,
the real periodicity of the outburst is
$P_{\rm average}=26.70$\,d (Sect. 1).
Because $P_{\rm average}$ is the average of $P_1$ and $P_2,$ one would expect,
when folding the data with  $P_{\rm average}$, a clustering  similar to or somehow intermediate to  what is obtained using
 $P_1$ and $P_2$ (shown in Fig.~\ref{beat}\,b). In contrast, Fig. \ref{beat}\,c reveals a
 double clustering.          
In Fig.  \ref{beat}\,c, we use a different colour for model data and observations  before or after the minimum
of the long-term modulation.
The GBI points before (red points) and after (green points)
50841 MJD   cluster separately with a shift of  0.5 in phase.
This effect is discussed in \citet{massijaron13} and \citet{jaronmassi13}.  
What we want to point out here is that
our model indeed reveals the same  double clustering  as  the GBI observations:
 model data  before (black points) and after (blue points)  50841 MJD
cluster separately with a shift of  0.5 in phase as well.

\begin{figure}[b]
\begin{center}
\includegraphics[width=7.cm,angle=0]{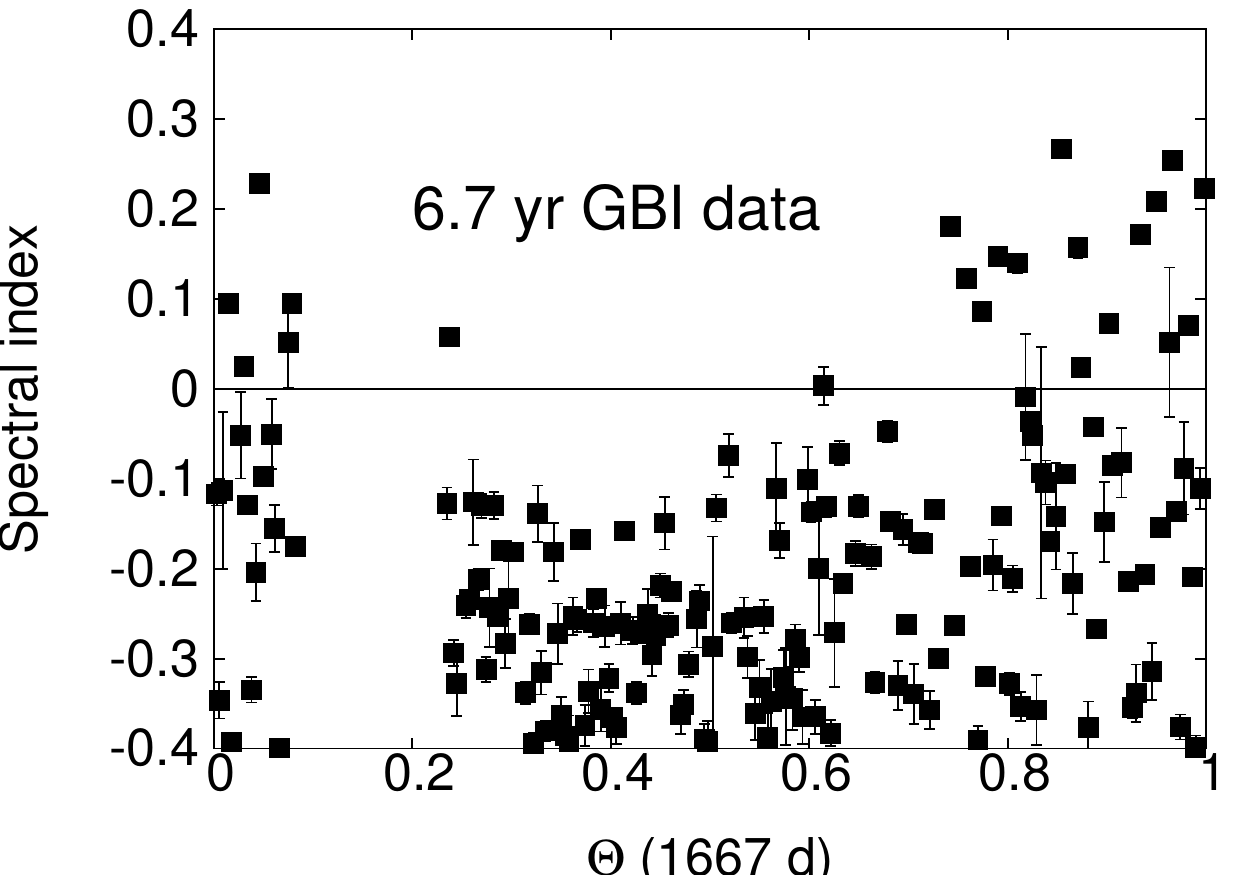}\\
\includegraphics[width=7.cm,angle=0]{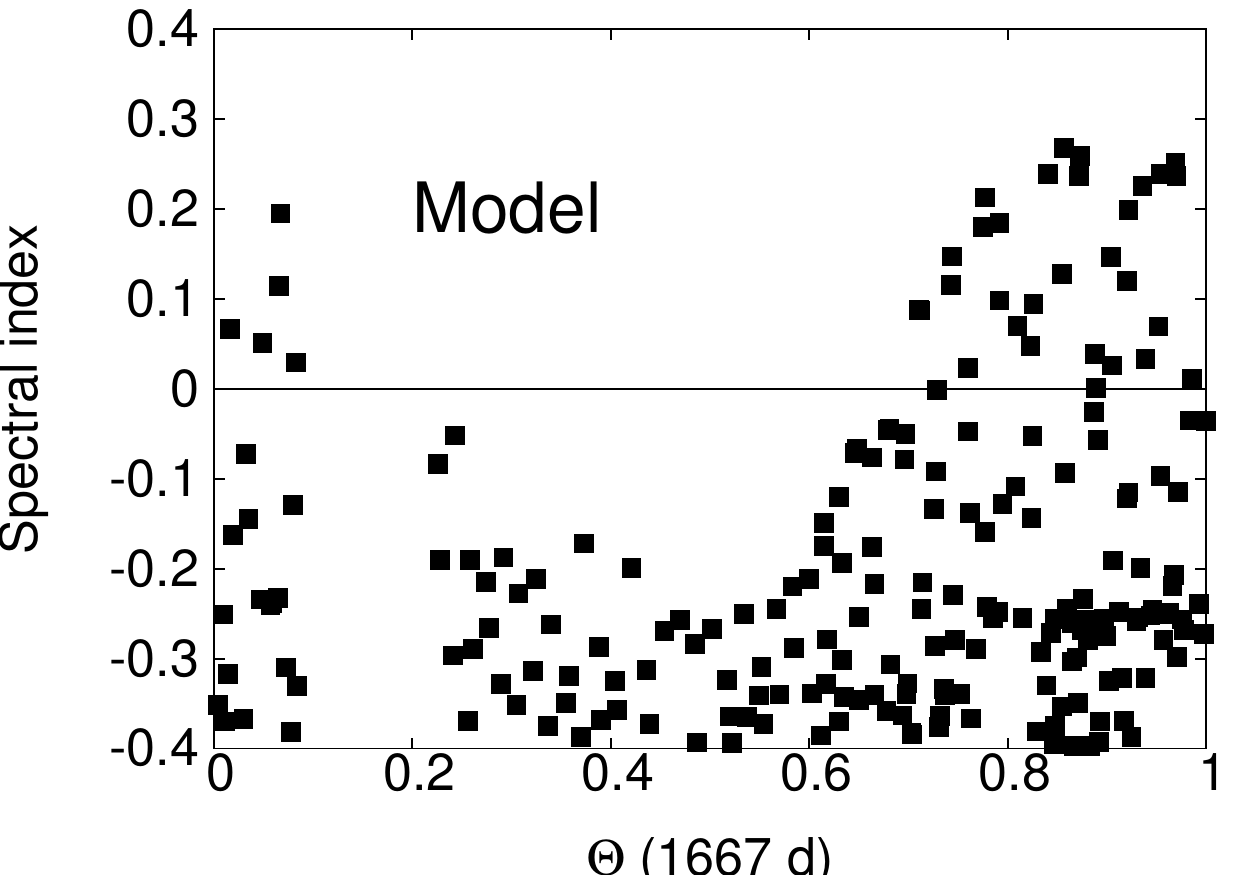}\\
\caption{ 
Top: Spectral index $\alpha$ folded with the long-term modulation (phase $\Theta$)
  from GBI data  at 2.2~GHz and 8.3~GHz  \citep{massikaufman09}.
Bottom: $\alpha$ vs $\Theta$  from model data  at 2.2~GHz and 8.3~GHz averaged over 4 days.}
\label{index}
\end{center}
\end{figure}

\subsection{Spectral index}
The top panel of Fig. \ref{index} shows  the observed spectral index,  from GBI data at 
$\nu= 2.2$~GHz and   8.3~GHz,  
folded  with the 1667\,d long-term modulation. 
The GBI data indicate that at some $\Theta$ phases, the emission in \lsi 
develops an optically thick spectrum ($\alpha \geq $ 0). 
This is the flat or inverted  spectrum typical of jets in microquasars as discussed in Sect. 2.1.
Our model of a precessing conical jet with a periodical increase in intrinsic 
emission   must therefore be able to reproduce  this spectral 
characteristic of \lsi  emission.

To make the comparison, we calculated the observed flux, $S_{\rm model}$, at 2.2 GHz just
 by replacing this  frequency in  equations shown in  Sect. 2 and  using the parameters obtained 
by fitting the model to GBI data at 8.3 GHz, i.e. the set of parameters used to generate all figures.
Indeed the model spectral index
shown at the bottom of  Fig. \ref{index}\, is consistent with the observed one  (Fig. \ref{index}\,top). In particular, we note that
the emission becomes optically thick 
at  $\Theta$s  where the maximum
emission of relativistic electrons occurs at  the smallest observing angles. In this case, in fact,
high values of the optical depth are obtained and 
the emission becomes optically thick (see Sect.2.3).  
A deep  analysis of the evolution of the opacity in the jet vs $\Phi$ and vs $\Theta$ 
is the argument of  a paper in preparation.
Whereas 
 the model presented here consists of
only a  conical jet, the following paper will  include the transient jet as well.
In microquasars,  the so-called transient jet always occurs after an  optically thick phase
(i.e., after the inverted/flat spectrum phase) and is associated to  a  large optically thin outburst 
\citep{fender04, massi11}.
In \lsi, the large optically thin outburst occurs about five days after the optical thick outburst
and  dominates at 2.2 GHz \citep{massikaufman09}.  
 The optically thin outburst suddenly reverses $\alpha$ from positive values  to the low value 
$\alpha\simeq - 0.5$, and this corresponds, as discussed in \citet[Sect. 5.1]{massikaufman09}, 
to a new population of energetic electrons with  an index power law  p=2, which is steeper 
than the one associated to the conical jet, modelled here, with p=1.8. 
It is rather interesting that existing  analyses of the origin of high energy emission in \lsi invoke
not  only two  populations of electrons, but also populations with  almost the  same 
index as results from radio analysis (p=1.8 for the conical jet and 
p=2 for the transient jet).
\citet{zabalza11} exclude GeV and TeV emission in \lsi being produced from
the same parent particle population and determine an index $\leq 1.8$ for Fermi data and $2.1$ 
for MAGIC data.

\subsection{VLBA maps vs model}
\begin{figure*}[]
\includegraphics[width=15.0cm,angle=0]{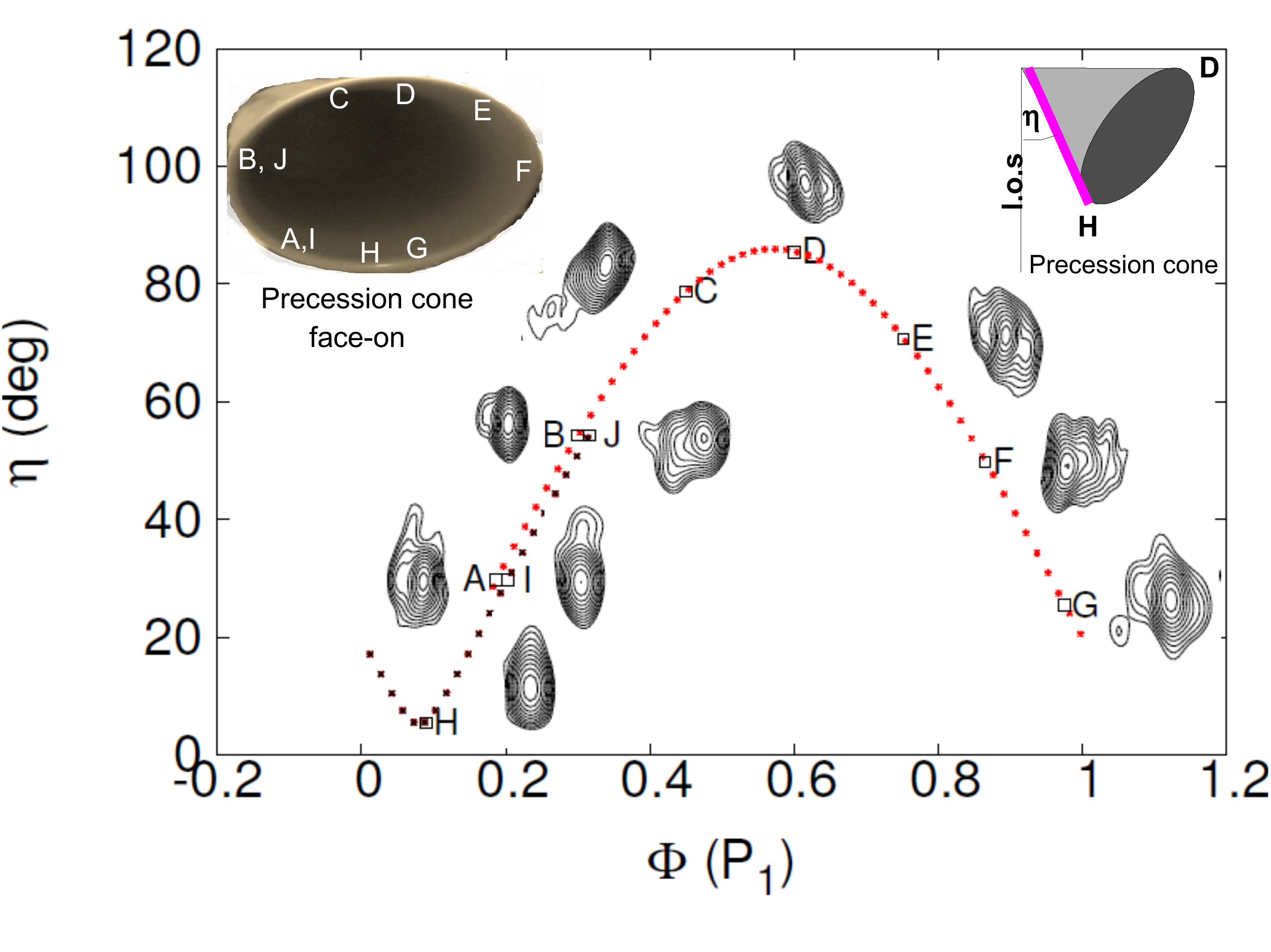}
\caption{
Angle $\eta$ between the jet axis and the line of sight 
at the epochs of the  VLBA runs A-J.
The trend of $\eta$ vs $\Phi (P_1)$ is shown together with the VLBA  maps \citep{massi12} (see Sect. 4.5). 
The VLBA runs cover 30 days, then more than one orbital cycle of 26.49 days;
runs I at $\Phi=0.203$ and J at $\Phi=0.316$ were  performed 
 one cycle later than runs A at $\Phi=0.187$  and B at $\Phi=0.30$.  
The shift of $\eta$ (corresponding to a shift of the maximum Doppler boosting)
is already appreciable in Fig. 7 even after just one $P_1$ cycle; i.e., the two 
$\eta$ curves (black and red ones)  vs $\Phi$ do not overlap.
}
\label{vlba}
\end{figure*}
Hints of a fast precessing jet in \lsi arose when 
two consecutive MERLIN images revealed a surprising variation
of 60$\degr$ in position angle in only one day (Massi et al. 2004).
A set of  VLBA observations, A-J, 
interlapsed  three days and  covering an interval of 30 days
were performed by \citet{dhawan06}. 
The VLBA images, about 5 mas in size (therefore more than 
a factor 10 the orbital size, $2 \times a = $0.45 mas)   were showing highly varying  features.
With our model we can determine the variation in the ejection angle 
 during the VLBA runs and compare these variations with the observed variations in position angle
of the maps.
The changes of $\eta$ are due to precession, but we put them
as usual as a function of the orbit, that is  of $\Phi(P_1)$, and relate them to changes in VLBA maps also given  as a function of $\Phi(P_1)$ in \citet{dhawan06} and \citet{massi12}.
The trend of $\eta$
vs $\Phi (P_1)$ is shown in Fig. \ref{vlba} together with the high dynamic range VLBA images by \citet{massi12}.

 In Fig.\ref{vlba} we see, as the plot of $\eta$ vs $\Phi$ reveals, that
 run H was performed at the minimum $\eta,$ and 
D was performed  nearly at the  maximum $\eta$.
These two extreme situations of $\eta$ for the two runs, H and D, are sketched in Fig.\ref{vlba} (right corner).
This important information   implies that 
two runs at similar $\eta$, but one performed   before and
the other after run D, 
refer to two jets pointing in opposite directions with respect to the axis 
of the precession cone (Fig.~7, left corner). 
If the model is correct, the position angle of the  associated radio structures  
should reflect this different orientation.
Indeed, the  structure for run E points towards the SW, whereas 
runs C and J show  structures pointing to SE.
Similarly, the structure at run B points E, whereas that for run F
points W.
Finally, for A,I,H,G, our model results in a  low $\eta$ angle, and this corresponds to a jet 
 pointing closest to the earth. One can see that the related VLBA structures 
develop indeed a N-S feature. 
In run A (G), a transition of the geometry between  B (F) and  H,I is even traceable. 
\subsection{Derived jet parameters}

The model fitting of Sect. 3 
determines the values of two normalization factors present in our model,  i.e. 
 the normalization factor for the jet emission
\beq 
 Q= {2 x_0^2 r_0 J_0 \over -(2-2~a_3-(p+2)a_2) D^2 }, 
\eneq
and 
 the normalization factor of the jet optical depth 
\beq
 q=(2.2~10^9)^{-(p+4)/2}{2 \chi_0r_0 \over -\kappa_0(2-2~a_3-(p+2)a_2)}
\eneq
where $r_0=x_0~\tan~\xi$.

Taking   expressions (12) and (13) into account
and assuming $p=1.8$ and  
$D=~2.3~$Kpc \citep{gregory79},
we can express the constant quantities $Q$ and $q$
in terms of the jet physical parameters 
\beq
Q= {x_0^3 tg \xi  ~  B_0^{1.4} A \over -(2-2~a_3-(p+2)a_2)}  4.4 \times 10^{-50}  
\eneq
\beq
q= {x_0 tg \xi~ B_0^{1.9} \over  -(2-2~a_3-(p+2)a_2)}  3.8 \times 10^{-3} 
\eneq
and derive, for the solution of Sect. 3, the values for the magnetic field intensity at the jet
basis, $B_0$, and that for  the normalization distance  $x_0$:
$$B_0 \simeq 9  ~~{\rm G} ~~~~~~~~~ x_0 \simeq  1.6~10^{13}  {\rm cm}.$$

An independent  estimate for the  magnetic field strength
is  the value
of the equipartition $B_{eq}\simeq 0.7~$G 
derived   
from VLBI data \citep{massi93} for  a source size $d\simeq 5.5~10^{13}$~cm, 
assuming equipartition between high energy particle energy and magnetic energy.
To compare the result of $B_0 \simeq 9 ~$G
for our model to  $B_{eq}$, we 
compute the jet  mean magnetic field $\tilde{B}$  
over its length $L$.
When considering the magnetic field functional form (Sect. 2.1),
the topology resulting in Sect. 3
and the value of $L$ (Sect. 2.3), from the expression
\beq
\tilde{B}={B_0 \over L} \int_1^{L} l^{-a_2}~dl, 
\eneq
we obtain 
$\tilde{B} \simeq 0.6~$G,  which is  consistent  with the value of $B_{eq}$
derived from the VLBI observation.

 From the parameters deduced above, 
 the   electron energy  content can  also be estimated.
 Since synchrotron emission   
 is centred on a peak spectral frequency $\nu$ 
\citep{ginzburgsyrovatski65} such as 
\beq
\nu = 1.8~10^6 \gamma ^2 B ,
\eneq
the electron energy,  $E= \gamma mc^2$,   can be derived for different values of the magnetic
field strength along the jet and for the two frequencies,  $\nu =2.2~10^9 Hz \div 8.3~10^9 Hz $, considered in this analysis.  At  the jet basis (l=1),
  $B=B_o$, and $\gamma \simeq 11 \div 22$ results.  For the jet end (l=L),  we must take both  the decreased magnetic field intensity and 
 relativistic electrons  adiabatic losses 
into account
(synchrotron losses are negligible)  so that an electron with $\gamma (l=1)$ at the jet basis 
arrives at the jet end with $\gamma(L)= \gamma (1) L^{-2/3}$ \citep{kaiser06}. To reproduce the observed emission for 
$\nu =2.2~10^9 Hz \div 8.3~10^9 Hz $ at the jet end, we need
\beq
\gamma(L) ^2 B(L)=\gamma(1) ^2 B_o  L^{-2/3-2}= \nu/ 1.8~10^6,
\eneq
i.e., for the solution of Sect.3, at least electrons with $\gamma (1)\simeq 371 \div 741 $.

In conclusion,  relativistic electrons   with  energies  in the range 
$10 ~mc^2 \la E \la 740~ mc^2$  are needed to reproduce
the  radio emission analysed in this paper.  Since the mean over the orbital period of  the adopted angular expression
(the one shown in Fig.~\ref{shift}\,top, see Sect.2.4 ) is  $ 0.58~  A$, the 
luminosity  injected in the two jets  in the form of relativistic electrons results
in\beq
L_{rel}= 2(\pi r_o^2c)~0.58 A \int^{E_{Max}}_{E_{min}} E^{1-p} dE  \simeq 5~10^{34}~~{\rm ergs/sec}.
\eneq

 This value is  only an indicative lower limit  for  the luminosity in the injected relativistic electron
 distribution since it is strictly related to the specific  part of the  radio spectrum
 analysed in this paper.
In particular the presence of synchrotron emission at higher  frequencies  
will imply  higher values for   electron energies  and,    consequently,  higher values  of $L_{rel}$.
Indeed, jets from X-ray binaries are supposed  to radiate from radio to 
optical wavelengths. The jet origin for the emission is supported by the correlation of the emission
in the different  bands \citep[and reference therein]{russell12}. 
In this respect, it is worth noting that in \lsi 
the timing analysis of optical data by \citet{zaitsevaborisov03} results in a period equal to $P_1$.
This result confirms what has previously been reported by \citet{mendelsonmazeh89, mendelsonmazeh94} 
and \citet{paredes94}.  
Particularly intriguing is that  the light curve  folded with this period shows 
that the  brightness increases
 in the orbital phase interval where the radio outburst occurs. 
Finally, as already pointed out in Sect. 4.4, 
the present analysis holds true for the conical jet with p=1.8 considered here.
Shocks in a transient jet may produce even more accelerated particles with a different
energy power-law distribution.

\subsection{The low/hard X-ray state}
The varying morphology of VLBA images, plus the lack of accretion disk
features (e.g., black body radiation) in the X-ray spectrum of \lsp, 
have led some authors to favour non-accretion models \citep[e.g.,][]{dubus13}.
VLBA images were discussed in Sect. 4.5, and  in this section we analyse  
the X-ray state of \lsp. In fact, an accreting object presents rather 
different X-ray states \citep{mcclintockremillard06}.  
While  the requested  black body component is the characteristic of  the thermal state
\citep{mcclintockremillard06}, 
radio and X-ray observations  of \lsi point towards a very low  low/hard  X-ray state.

As shown by \citet{corbel13},  in  the early phase of  jet formation,
the  low density of particles in the jet  produces an optically thin 
synchrotron power-law spectrum; as the density of the jet plasma 
increases, this leads to a transition to higher optical
depth and to a flat/inverted radio spectrum ($\alpha \geq 0$).
When sources having these  self-absorbed compact radio jets are observed in 
X-rays, they reveal a low/hard X-ray state \citep{gallo03,fender04}. 
Concerning \lsp,  it is known that its  persistent radio emission  
 develops a  flat/inverted radio spectrum at some orbital phases 
(exactly where an increased accretion is predicted) 
as shown by GBI observations at  2.2 GHz and 8.3 GHz
\citep{massikaufman09}. In addition,  a flat spectrum has been recently observed by 
the Effelsberg 100-m telescope 
at four frequencies, from 4.85~GHz to 14.6~GHz (Fig. 4.13 \citet{zimmermann13}).
 
Modelling of the low/hard state, especially at very low luminosities, 
favours an accretion  disk
truncated at a large radius
 \citep[and references therein]{fender01, gardnerdone13}.
The origin of the X-ray power-law emission
with a typical photon index $\Gamma \sim 1.7$
  \citep{remillardmcclintock06} 
  (for \lsi in the range $1.5-1.8$, \citet{sidoli06, chernyakova06, zimmermannmassi12})            
is still controversial, if
it is due to a  Comptonizing corona or/and  to the jet \citep{markoff01}.
Nevertheless,  it is well established that this X-ray power-law emission
that characterizes the low/hard state is correlated with the radio emission from the self-absorbed jet
($L_R  \propto L_X^{0.7}$,  \citet{gallo03}).
 The luminosity values for \lsi are 
 $L_X=(1-6)10^{34}$ ergs/sec
and    $L_R=(1-17)\times 10^{31}$ ergs/sec \citep{combi04}.
From  Eqs. 3 and 5 in \citet{merloni03}, we obtain 
\beq
\xi_{RX}= {{\rm log} L_R -\xi_{{\rm RM}} {\rm log} M - b_{\rm R} \over {\rm log} L_X}
\eneq
where  $\xi_{{\rm RM}}=0.69-0.89$ and  $b_{\rm R}=3.26-11.38$ 
\citep{merloni03}.
For a mass of 3 $M_\odot$ 
(where the influence of the  mass difference between a neutron star 
of 1.4 $M_\odot$ and
a stellar mass black hole of $3 - 15 M_\odot$ is negligible in the result),
we obtain  $\xi_{RX}~=0.55-0.85$. 
This value is  consistent with the 
solutions  $\xi_{RX}=0.49-0.71$ by \citet{merloni03}
and with the value of $\sim 0.7$ by \citet{gallo03}
for accreting black holes in low/hard state,
and it is clearly below the value $> 1.4$ 
 given by \citet{migliarifender06} for accreting  neutron
stars in the low/hard state.  
Even though the nature of the compact object in \lsi is still unknown
\citep{casares05},  the  agreement with the low/hard state of black
hole X-ray binaries would indeed  favour the black hole scenario.

The luminosity for accreting black holes
in  the low/hard 
state is  about  $L_{\rm X} \sim 10^{36}~ {\rm erg/sec}$
 that   may drop to
 $L_{\rm X}= 10^{30.5} - 10^{33.5}$ erg/s \citep{mcclintockremillard06}
 at the  lowest  phase, which is  called the quiescent state.
  With its luminosity $L_{\rm X}=(1-6)10^{34}$ ergs/sec, \lsi 
is clearly  in a very low  low/hard  X-ray state. 
 That is  a similar case to MWC 656,
the recently discovered  Be-black hole system, which is in the quiescent state                    
\citep{casares14}.
During  the low-hard state  the liberated energy of the accretion
is thought to be converted into magnetic energy that  powers the relativistic jet 
typically observed in this  state \citep[and references therein]{smith12}. 
In a system in quiescence, \citet{gallo06} demonstrate that the outflow kinetic
power accounts for a sizable fraction of the accretion energy budget.

\section{Conclusions}

In this paper we have developed a model that reproduces the observed long-term modulation
of the   radio
emission of the binary system \lsi. The hypothesis is that the radio emission is due to a  
jet precessing with a period of 
$P_2=\unit[26.92 \pm 0.07]{days}$.
The jet is  anchored on the accretion disk of a compact object orbiting, with a period of 
$P_1=\unit[26.49 \pm 0.07]{days}$, the Be star. 
Variable accretion around the eccentric orbit increases the  relativistic electron density in the jet
at a particular orbital phase.
We fitted  our theoretically derived  flux density, $S_{\rm model}(t)$  (see Eq. 1)  to 6.5 yr GBI radio observations
and determined  the parameters of Table 1.  We found that  
the  precessing ($P_2$) jet model with intrinsic periodical variations ($P_1$) of emitting particles
 reproduces all seven  observational facts:  
\begin{enumerate}
\item
the  long-term modulation  \citep[and references there]{gregory02}, see Fig. \ref{mod}\,a and Fig. \ref{shift}\,bottom; 
\item 
the  orbital shift of the peak of the radio outburst \citep{paredes90}, see Fig. \ref{shift}\,bottom;
\item
the periodical outburst with orbital occurrence \citep[and references there]{gregory02}, see Figs. \ref{mod}\,b and c and  Fig. \ref{beat}\,a; 
\item 
        the   double clustering for data folded with   
        $P_1+P_2\over 2$  \citep{massijaron13}, see Fig. \ref{beat}\,c;
\item 
the  spectral index evolution \citep{massikaufman09}, see Fig. \ref{index};
\item 
a variation of the ejection angle suggestive of 
the  rapid rotation in position angle of VLBA maps \citep{dhawan06,massi12}, see Fig. \ref{vlba};  
\item
the strength of the magnetic field \citep{massi93}, see Sect. 4.6.
\end{enumerate}

The maximum of the long-term modulation occurs  
when the maximum ejection of relativistic electrons takes place  and  at  the same time 
the  approaching jet is forming  the smallest possible angle with the line of sight
(i.e., the Doppler boosting is at its maximum).
After a cycle of duration $P_2$,  
when the approaching jet again forms the smallest angle $\eta$, 
the  ejection
is now already on its decay phase (being $P_1 < P_2$) and its Doppler boosting gives rise to
an outburst with a lower amplitude than that of one cycle before. 
As we quantitatively show in 
Sect. 4.2, the peak of the outburst gets lower and lower
at each cycle 
 until the most unfavorable situation occurs, i.e., when the jet is
pointing towards us while the main ejection is at its minimum.
After this point, it happens that 
when the approaching jet again forms the smallest $\eta$ angle, there is  
the beginning of a new ejection of particles and the observed flux density slowly rises again.
The period of 1667$\pm$8 days \citep{gregory02} of the long-term modulation is then just the number of $P_1$
cycles needed to compensate for the difference $P_2-P_1$ and synchronize $P_1$  again to  $P_2$,
which is just $[P_2/(P_2-P_1)]P_1$ \citep{massijaron13}.

Our  model,   meant in the first instance  to explain the long-term modulation affecting
the radio emission of \lsp, can become a powerful tool for understanding the physical
processes at work in this gamma-ray binary. 
For this purpose the model should be   extended to include  
internal shocks in the conical flow, i.e., the transient jet.
As  discussed in Sect. 4.4. the transient jet observed in \lsi is associated to a   
second  population of relativistic electrons, with $p=2.0$, which is therefore different from
the population with $p=1.8$,  forming the conical jet analysed in this paper.
It is quite interesting that two similar  populations 
are invoked to explain emission in the  TeV  and GeV band  \citep{zabalza11}.
The important implications, if confirmed, are that the GeV emission is related to inverse Compton
of the electrons of the conical jet and that the TeV emission is related to  electrons of the transient jet.

This model, with conical jet and transient jet can be tested by fitting it  to 
the radio spectral  index data analysed in \citet{massikaufman09}. 
The  theory of accretion in  an eccentric orbit  (Sect. 2.4) also predicts an ejection around
periastron passage,
along with the ejection at orbital phase
$\Phi$=0.6 considered here  (i.e., towards apoastron).
By fitting the model to the  radio spectral  index data, one could  verify 
if  each of the two populations of electrons is  generated
twice along the orbit, i.e., that the transient jet occurs at periastron as well.   

Indeed, at  different epochs,
emission towards  apoastron \citep{albert09} and  periastron
 \citep{acciari11} was detected by Cherenkov telescopes.
Different  Doppler boosting 
 of the two ejections along the eccentric orbit   
could   explain the observed variations.
Maximum Doppler boosting occurring between the two ejections 
could explain
the broad gamma-ray emission without any clear orbital modulation observed  
in Fermi-LAT data \citep{hadasch12, ackermann13}.

Finally, considering that one contribution to the gamma-ray emission consists of upscattered stellar photons,
the role of Doppler boosting in  gamma-ray emission 
is probably  even  more important than in the radio band. 
In fact,  the amplification due to the Doppler factor for external inverse 
Compton is higher than for synchrotron emission
\citep[and reference therein]{kaufman02}.  
\acknowledgements
 We thank Brenda Burton, Lars Fuhrmann, Fr\'ed\'eric Jaron, J\"urgen Neidhofer, and Lisa 
Zimmermann for interesting discussions 
and the anonymous referee for the careful reading of our 
 manuscript and the useful comments made for its improvement.
The Very Long Baseline Array is operated by the USA National Radio Astronomy 
Observatory, which is a facility of the USA National Science Foundation operated under cooperative agreement by Associated Universities, Inc.
The Green Bank Interferometer is a facility of the National
Science Foundation operated by the NRAO in support of NASA High Energy
Astrophysics programmes.
MINUIT  is the Function Minimization and Error Analysis
CERN Program Library. We thank Dirk Muders for technical support.

\bibliographystyle{aa}

\begin{thebibliography}{}
\bibitem[Acciari et al.(2011)]{acciari11} Acciari, V.~A., Aliu, 
E., Arlen, T., et al.\ 2011, \apj, 738, 3 
\bibitem[Ackermann et al.(2013)]{ackermann13} Ackermann, M., 
Ajello, M., Ballet, J., et al.\ 2013, \apjl, 773, L35 
\bibitem[Albert et al.(2009)]{albert09} Albert, J., Aliu, E., 
Anderhub, H., et al.\ 2009, \apj, 693, 303 
\bibitem[Bailyn et al.(1995)]{bailyn95} Bailyn, C.~D., Orosz, 
J.~A., McClintock, J.~E., \& Remillard, R.~A.\ 1995, \nat, 378, 157 
\bibitem[Blandford \& K\"onigl(1979)]{blandfordkoenigl79} 
Blandford, R.~D., \&  K\"onigl, A.\ 1979, \apj, 232, 34 
\bibitem[Bondi \& Hoyle(1944)]{bondihoyle44} Bondi, H., \& Hoyle, F.\ 1944, \mnras, 104, 273 
\bibitem[Bosch-Ramon et al.(2006)]{boschramon06} Bosch-Ramon, V., Paredes, J.~M., Romero, G.~E., \& Rib{\'o}, M.\ 2006, \aap, 459, L25 
\bibitem[Casares et al.(2005)]{casares05} Casares, J., Ribas, I., 
Paredes, J.~M., Mart{\'{\i}}, J., 
\& Allende Prieto, C.\ 2005, \mnras, 360, 1105 
\bibitem[Casares et al.(2014)]{casares14} Casares, J.,                              
Negueruela, I., Ribo, M., et al.\ 2014, Nature, 505, 378
\bibitem[Chernyakova et al.(2006)]{chernyakova06} Chernyakova, M., 
Neronov, A., \& Walter, R.\ 2006, \mnras, 372, 1585 
\bibitem[Combi et al.(2004)]{combi04} 
Combi, J.~A., Rib{\'o}, M., Mirabel, I.~F., \& Sugizaki, M.\ 2004, \aap, 422, 1031 
\bibitem[Corbel et al.(2013)]{corbel13} Corbel, S., Aussel, H., 
        Broderick, J.~W., et al.\ 2013, \mnras, 431, L107 
\bibitem[Dhawan et al.(2006)]{dhawan06} Dhawan, V.,  Mioduszewski, A., \&
Rupen, M. 2006, Proceedings of  the VI Microquasar Workshop, p. 52.1
\bibitem[Dubus(2013)]{dubus13} Dubus, G.\ 2013, \aapr, 21, 64 
\bibitem[Fender(2001)]{fender01} Fender, R.~P.\ 2001, \mnras, 
322, 31 
\bibitem[Fender et al.(2004)]{fender04} Fender, R.~P., Belloni, 
T.~M., \& Gallo, E.\ 2004, \mnras, 355, 1105 
\bibitem[Gallo et al.(2003)]{gallo03} Gallo, E., Fender, R.~P., 
\& Pooley, G.~G.\ 2003, \mnras, 344, 60 
\bibitem[Gallo et al.(2006)]{gallo06} Gallo, E., Fender, R.~P., 
        Miller-Jones, J.~C.~A., et al.\ 2006, \mnras, 370, 1351 
\bibitem[Gregory et al.(1979)]{gregory79} Gregory, P.~C., Taylor, 
A.~R., Crampton, D., et al.\ 1979, \aj, 84, 1030 
\bibitem[Gardner \& Done(2013)]{gardnerdone13} 
Gardner, E., \& Done, C.\ 2013, \mnras, 434, 3454 
\bibitem[Ginzburg \& Syrovatski(1965)]{ginzburgsyrovatski65}
 Ginzburg, V. L., \& Syrovatski, S. I. 1965 ARA\&A, 3, 297 
\bibitem[Gregory et al.(1999)]{gregory99} Gregory, P.~C., 
Peracaula, M., \& Taylor, A.~R.\ 1999, \apj, 520, 376 
\bibitem[Gregory(2002)]{gregory02} Gregory, P.~C. 2002, ApJ, 575, 427
\bibitem[Gregory \& Neish(2002)]{gregoryneish02} 
Gregory, P.~C., \& Neish, C.\ 2002, \apj, 580, 1133 
\bibitem[Grundstrom et al.(2007)]{grundstrom07} Grundstrom, E.~D., 
Caballero-Nieves, S.~M., Gies, D.~R., et al.\ 2007, \apj, 656, 437 
\bibitem[Hadasch et al.(2012)]{hadasch12} Hadasch, D., Torres, 
D.~F., Tanaka, T., et al.\ 2012, \apj, 749, 54 
\bibitem[Hjellming \& Rupen(1995)]{hjellmingrupen95} 
Hjellming, R.~M., \& Rupen, M.~P.\ 1995, \nat, 375, 464 
\bibitem[Jaron \& Massi(2013)]{jaronmassi13}  Jaron, F., \&  Massi, M., \ 2013, \aap, 
accepted
\bibitem[Kaiser(2006)]{kaiser06} Kaiser, C.~R.\ 2006, \mnras, 
367, 1083
\bibitem[Kaufman Bernad{\'o} et 
al.(2002)]{kaufman02} Kaufman Bernad{\'o}, M.~M., Romero, G.~E., \& Mirabel, I.~F.\ 2002, \aap, 385, L10 
\bibitem[Larwood(1998)]{larwood98} Larwood, J.\ 1998, \mnras, 299, L32
\bibitem[Longair (1994)]{longair}Longair, M.S. ,1981, ''High energy astrophysics'', Cambridge University Press
\bibitem[Markoff et al.(2001)]{markoff01} Markoff, S., Falcke, H.,
\& Fender, R.\ 2001, \aap, 372, L25
\bibitem[Marti \& Paredes(1995)]{martiparedes95} Marti, J., \& Paredes,
J.~M.\ 1995, \aap, 298, 151
\bibitem[Martin et al.(2008)]{martin08} Martin, R.~G., Tout, 
C.~A., \& Pringle, J.~E.\ 2008, \mnras, 387, 188 
\bibitem[Massi et al.(1993)]{massi93} Massi, M., Paredes, J.~M., Estalella, R., \& Felli, M.\ 1993, \aap, 269, 249 
\bibitem[Massi et al.(2004)]{massi04} Massi, M., Rib\'o, M,  Paredes, J. M., et al. 2004, A\&A, 414, L1
\bibitem[Massi \& Kaufman Bernad{\'o}(2009)]{massikaufman09} Massi, M., \& Kaufman Bernad{\'o}, M.\ 2009, \apj, 702, 1179 
\bibitem[Massi \& Zimmermann(2010)]{massizimmermann10} Massi, M., \&
Zimmermann, L. \ 2010, \aap, 515, A82
\bibitem[Massi(2011)]{massi11} Massi, M.\ 2011, \memsai, 82, 24 
\bibitem[Massi et al.(2012)]{massi12} Massi, M., Ros, E., \& Zimmermann, L.\ 2012, \aap, 540, A142 
\bibitem[Massi 
\& Jaron(2013)]{massijaron13} Massi, M., \& Jaron, F.\ 2013, \aap, 554, A105 
\bibitem[McClintock \& Remillard(2006)]{mcclintockremillard06}
McClintock, J.~E., \& Remillard, R.~A.\ 2006, \textit{Compact stellar X-ray sources}, Cambridge University Press, p. 157
\bibitem[{Mendelson \& Mazeh(1989)}]{mendelsonmazeh89}
Mendelson, H., \& Mazeh, T. 1989, MNRAS, 239, 733
\bibitem[{Mendelson \& Mazeh(1994)}]{mendelsonmazeh94}
Mendelson, H., \& Mazeh, T. 1994, MNRAS, 267, 1
\bibitem[Merloni et al.(2003)]{merloni03} Merloni, A., Heinz, S.,
\& di Matteo, T.\ 2003, \mnras, 345, 1057
\bibitem[Migliari 
\& Fender(2006)]{migliarifender06} Migliari, S., \& Fender, R.~P.\ 2006,
\mnras, 366, 79 
\bibitem[Mirabel \& Rodr{\'{\i}}guez(1999)]{mirabelrodriguez99} Mirabel, I.~F., \& Rodr{\'{\i}}guez, L.~F.\ 1999, \araa, 37, 409 
\bibitem[Nelder \& Mead(1965)]{neldermead65}
Nelder, J.A., \& Mead, R.\ 1965, 
Comput. J., 7, 308 
\bibitem[Paredes et al.(1990)]{paredes90} Paredes, J.~M., Estalella, R., \& Rius, A.\ 1990, \aap, 232, 377 
\bibitem[Paredes et al.(1994)]{paredes94} Paredes, J.~M., et al.\ 1994, \aap, 288, 519
\bibitem[Potter 
\& Cotter(2012)]{pottercotter12} Potter, W.~J., \& Cotter, G.\ 2012, \mnras, 423, 756 
\bibitem[Remillard \& McClintock(2006)]{remillardmcclintock06}
Remillard, R.~A., \& McClintock, J.~E. 2006, \araa, 44, 49
\bibitem[Romero et al.(2007)]{romero07} Romero, G.~E., Okazaki, A.~T., Orellana, M., \& Owocki, S.~P.\ 2007, \aap, 474, 15
\bibitem[Russell et al.(2013)]{russell12} Russell, D.~M., 
Markoff, S., Casella, P., et al.\ 2013, \mnras, 429, 815 
\bibitem[Smart \& Green(1977)]{smartgreen77} Smart, W.~M., \& Green, E.~b.~R.~M.\ 1977, Textbook on Spherical Astronomy, by W.~M.~Smart, Edited by R.~M.~Green, Cambridge, UK: Cambridge University Press, 1977.
\bibitem[Sidoli et al.(2006)]{sidoli06} Sidoli, L., Pellizzoni, A., Vercellone, S., et al.\ 2006, \aap, 459, 901 
\bibitem[Smith(2012)]{smith12} Smith, M.~D.\ 2012, 
        Astrophysical Jets and Beams,  Cambridge University Press, 2012
\bibitem[Taylor et al.(1992)]{taylor92} Taylor, A.~R., Kenny, H.~T.,
Spencer, R.~E., \& Tzioumis, A. 1992, \apj, 395, 268
\bibitem[Urry 
\& Padovani(1995)]{urrypadovani95} Urry, C.~M., \& Padovani, P.\ 1995, \pasp, 107, 803
\bibitem[Zabalza et al.(2011)]{zabalza11} Zabalza, V., Paredes, J.~M., \& Bosch-Ramon, V.\ 2011, \aap, 527, A9 
\bibitem[Zimmermann \& Massi(2012)]{zimmermannmassi12} Zimmermann, L., \& Massi, M.\ 2012, \aap, 537, A82 
\bibitem[Zaitseva \& Borisov(2003)]{zaitsevaborisov03} Zaitseva, G.~V., \& Borisov, G.~V.\ 2003, Astronomy Letters, 29, 188
\bibitem[Zimmermann(2013)]{zimmermann13}
Lisa Zimmermann: Variability of radio and TeV emitting X-ray binary systems
 The case of \lsp. Bonn, Univ., Diss., 2013
URN:urn:nbn:de:hbz:5n-33175
\end{thebibliography}

\begin{appendix}
        \section{Optical depth through a stratified jet}

 The integration  upper limit  in the  intensity expression in Eq.(7), $\tau_{end}(l)$,  represents the jet's 
 maximum optical depth. 
Figure A1
shows a cut of the jet structure where the
 line of sight makes an angle $\eta$ with the jet axis, $l$.  When looking at the figure it is easy to understand
 how  photons  coming out  from the jet side facing
 the observer at  generic distance $x_0 l$ originate  along the integration path inside the jet, i.e., along  the segment AB for the case of the approaching jet shown in the figure.
 \begin{figure}[h]
\includegraphics[width=8.0cm,angle=0]{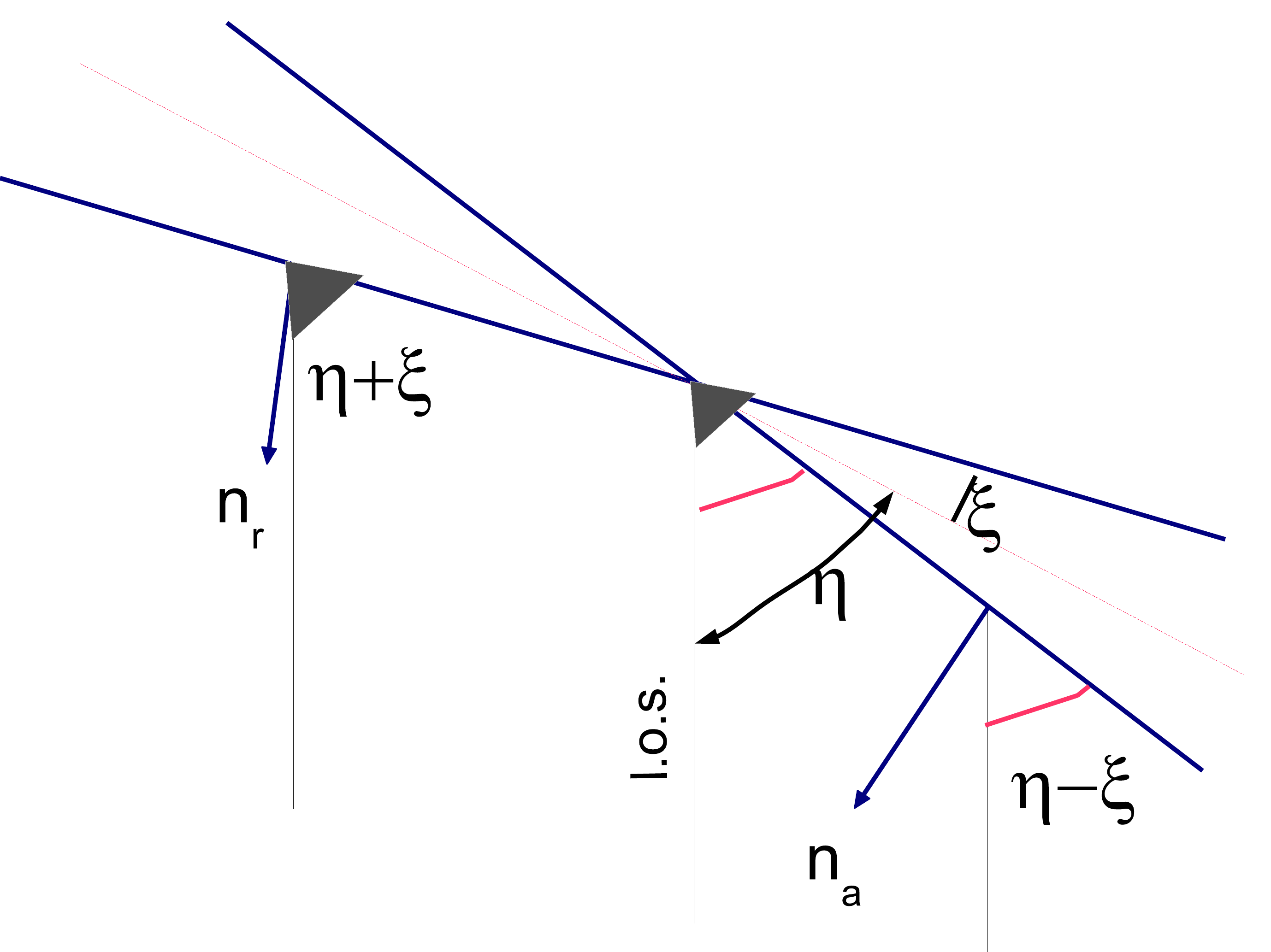}\\
\includegraphics[width=9.0cm,angle=0]{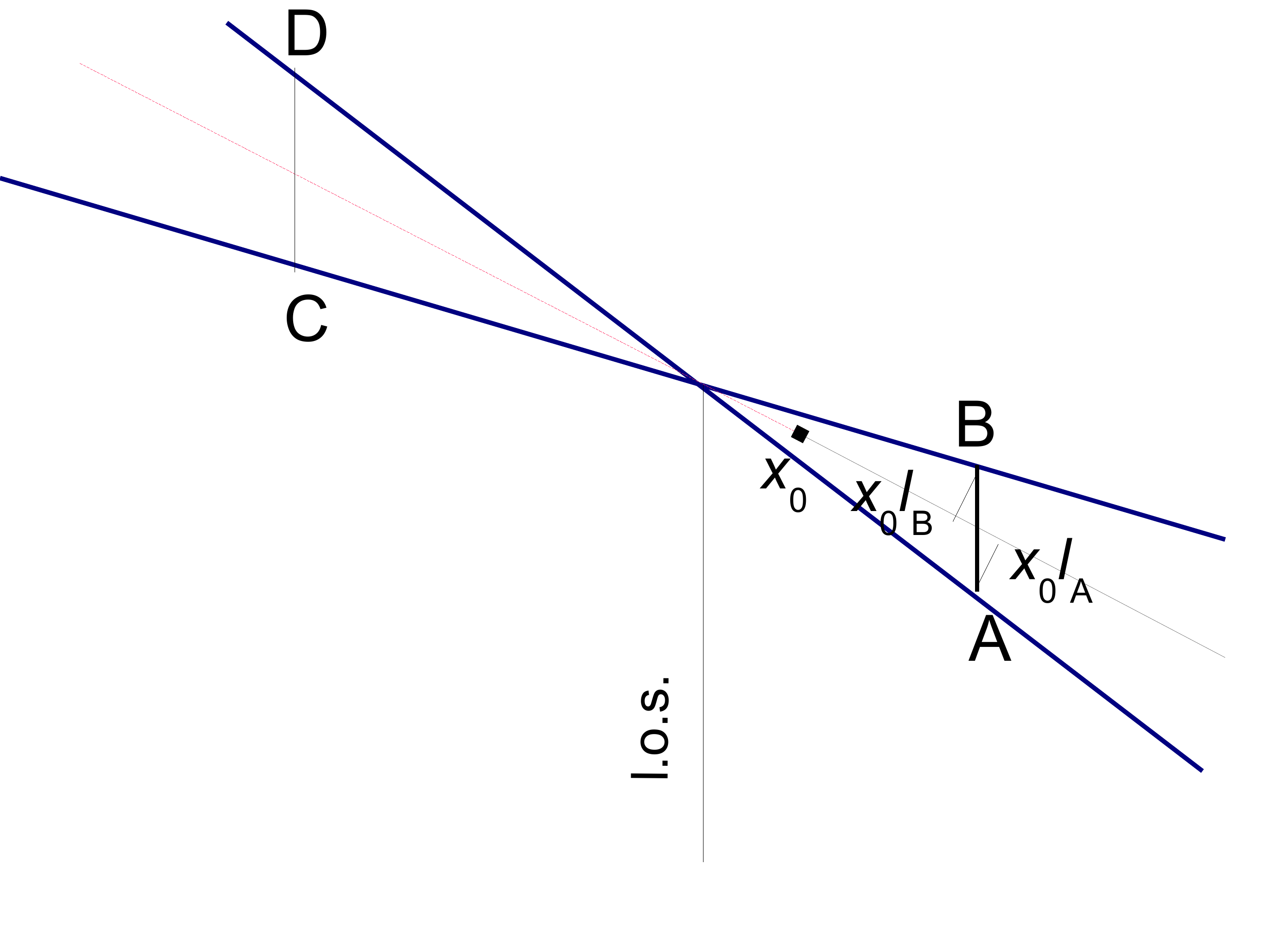}
\caption{Top: Perpendicular directions (n) to the surfaces facing the line of sight 
(l.o.s.) of the approaching and receding components of a 
jet with aperture $\xi$  and an angle $\eta$ with respect to the line of sight.
Bottom: Optical path  through the two jet components (see text).} 
\label{jet_geometry} 
\end{figure}

 If we call $l_{A}$ the value of the axial distance corresponding to point $A$ and $l_{B}$ that corresponding to point $B$, the following relationships describe the geometrical projection of the optical path $AB$ on the
 jet axis and on the jet radius, respectively: 
 
\beq
AB~\cos~\eta=x_0~(l_{A}-l_B)  
\eneq
 
\beq
AB~\sin~\eta=r(l_A)+r(l_B)=r_0~(l_A+l_B)        
\eneq

The jet region included between $l_A$ and $l_B$, i.e., $l_A \ge l \ge l_B$,  is the one
that  contributes to the emission coming out of the jet at $l=l_A$. It is evident from the figure that
the emitting region is different for different values of $l_A$.  To quantify  the depth of
the involved region, 
the value of  $l_B$  can easily be derived from the above expressions  and results in
\beq
l_B=l_A~ {\tan~\eta-\tan~\xi \over \tan~\eta+\tan~\xi }
\eneq
All the above relationships  hold in the limit that the angle  $\eta$, between the jet axis and the line of sight, is larger than the jet opening angle $\xi =r_0/x_o$; i.e.,  $\tan~\eta>\tan~\xi$. 

Once $l_B$ is known following Eq.(8),  the optical depth at the distance $l_A$  is 
\beq
\tau_A- \tau_B = -  \int^{l_A}_{l_B}  \chi _{\nu} ~ {\rm d}x = ~ \tau_0 l_A^{~C}~[1-(l_B/l_A)^{~C}]
\eneq
where  $C=1-a_3-(p+2)a_2/2      $ and  $\tau_A=0$ 
assuming  no emission or absorbtion   between the
observer and the jet.

Since $l_A$ represents our generic distance along the jet axis, we can define for the approaching jet the optical depth $\tau_{end} \equiv \tau_B$ as 
\beq
[\tau_{\rm end}]_{\rm app}(l)=\tau_0 l^{~C}~\large \left [\large \left({\tan~\eta-\tan~\xi \over \tan~\eta+\tan~\xi }\large \right)^{~C}-1\large \right].
\eneq
For the receding jet the same procedure can be performed by paying attention to 
some different signs (note that also for the receding jet the positive direction for $l$ is defined going away from the core), to arrive 
at
\beq
[\tau_{\rm end}]_{\rm rec}(l)=\tau_0 l^{~C}~\large \left [1-\large \left({\tan~\eta+\tan~\xi \over \tan~\eta-\tan~\xi }\large \right)^{~C}\large \right].
\eneq

\end{appendix}
\end{document}